\documentclass[aps,prr,twocolumn,superscriptaddress,amsmath,amssymb]{revtex4-2}
\usepackage[colorlinks=true,linkcolor=blue,anchorcolor=red,citecolor=blue, urlcolor=blue]{hyperref}
\usepackage{graphicx} 
\usepackage{dcolumn} 
\usepackage{bm} 
\usepackage{color}  

\begin{document}

\title{Magnetic topological transistor exploiting layer-selective transport} 

	\author{Hai-Peng Sun}
	\email{haipeng.sun@physik.uni-wuerzburg.de}
	\affiliation{\mbox{Institute for Theoretical Physics and Astrophysics, University of W\"urzburg, W\"urzburg 97074, Germany}}

	\author{Chang-An Li}
	\email{changan.li@uni-wuerzburg.de}
	\affiliation{\mbox{Institute for Theoretical Physics and Astrophysics, University of W\"urzburg, W\"urzburg 97074, Germany}}

	\author{Sang-Jun Choi}
	\affiliation{\mbox{Institute for Theoretical Physics and Astrophysics, University of W\"urzburg, W\"urzburg 97074, Germany}}
	
	\author{Song-Bo Zhang}
	\affiliation{Department of Physics, University of Z\"urich, Winterthurerstrasse 190, Z\"urich 8057, Switzerland}

	\author{Hai-Zhou Lu}
	\affiliation{Shenzhen Institute for Quantum Science and Engineering and Department of Physics, Southern University of Science and Technology (SUSTech), Shenzhen 518055, China}
	\affiliation{Shenzhen Key Laboratory of Quantum Science and Engineering, Shenzhen 518055, China}
	
	\author{Bj\"orn Trauzettel}
	\affiliation{\mbox{Institute for Theoretical Physics and Astrophysics, University of W\"urzburg, W\"urzburg 97074, Germany}}
	\affiliation{W\"urzburg-Dresden Cluster of Excellence ct.qmat, Germany}

\date{\today }

\begin{abstract}
We propose a magnetic topological transistor based on MnBi$_{2}$Te$_{4}$, in which the ``on" state (quantized conductance) and the ``off" state (zero conductance) can be easily switched by changing the relative direction of two adjacent electric fields (parallel vs. antiparallel) applied within a two-terminal junction. We explain that the proposed magnetic topological transistor relies on a novel mechanism due to the interplay of topology, magnetism, and layer degrees of freedom in MnBi$_{2}$Te$_{4}$. Its performance depends substantially on film thickness and type of magnetic order. We show that ``on" and ``off" states of the transistor are robust against disorder due to the topological nature of the surface states. 
Our work opens an avenue for applications of layer-selective transport based on the topological van der Waals antiferromagnet MnBi$_{2}$Te$_{4}$.
\end{abstract}

\maketitle

\section{Introduction}
Topological insulators (TIs) are bulk insulators with topological Dirac surface states protected by time-reversal symmetry \cite{Moore10NBirth,Hasan10RMPColloquium,Qi11RMPTopological}. Soon after the realization of TIs, various efforts have been made to exploit the role of topology in fabricating field-effect transistors~\cite{Xue11NNTopological,Xiu11NNManipulating}. Breaking time-reversal symmetry by magnetic doping in TIs, this opens a sizable gap in the spectrum of the 
Dirac surface states. This gap allows for the realization of topological transistors \cite{Wray12NPTopological, Checkelsky12NPDiracfermionmediated, Yu10SQuantized}, which could be elementary building blocks in topological electronics. Topological transistors may also be realized by exploiting topological phase transitions~\cite{Michetti13APLDevices,Qian14SQuantum,Liu14NMSpinfiltered,Wang15PRLElectrically,Liu15NLSwitching,Collins18NElectricfieldtuned}.  

Recently, the intrinsic antiferromagnetic TI MnBi$_2$Te$_4$ has been discovered \cite{Zhang19PRLTopological, Otrokov19NPrediction, Rienks19NLarge, Gong19CPLExperimental,Li19SAIntrinsic,Otrokov19PRLUnique,Sun19PRLRational,Lei20PNASMagnetized,Wang20PRBDynamical,Zhang20PRLMobius,Lian20PRLFlat,Lei21PRMGatetunable,Wei21PRBRenormalization,Varnava21NCControllable,Gu21NCSpectral,Li21PRLCoexistence,Chen21PRBUsing,Otrokov19NPrediction,Chen21PRBKoopmans,Du20PRRBerry,Liu20PRBAnisotropic,Liu21Dissipative,Cai22NCElectric}, which exhibits large magnetic surface gaps ($\sim70$ meV) \cite{Otrokov19NPrediction} and high mobilities (\textgreater 1000 cm$^2$/Vs)  \cite{Gao21NLayer,Liu21NCMagneticfieldinduced}. While extensive research efforts have been devoted to the topological electronic structure of the MnBi$_2$Te$_4$ family~\cite{Gong19CPLExperimental, Otrokov19NPrediction,  Rienks19NLarge,Chen19NCIntrinsic,Hao19PRXGapless,Li19PRXDirac,Chen19PRXTopological,Swatek20PRBGapless,Wu20PRXDistinct,Lu21PRXHalfMagnetic,Vidal21PRLOrbital,Lee21PRXEvidence}, their transport properties remain largely unexplored despite some magnetotransport measurements~\cite{Deng20SQuantum, Liu20NMRobust,Deng21NPHightemperature,Liu21NCMagneticfieldinduced,Ge20NSRHighChernnumber,Gao21NLayer,Ovchinnikov21NLIntertwined,Ge22PRBMagnetizationtuned,Liang22NLApproaching,Cai22NCElectric,Lei22PRBMagnetically}. Recent studies have revealed that electric fields work as convenient knobs to control the transport properties of MnBi$_{2}$Te$_{4}$  \cite{Du20PRRBerry,Gao21NLayer,Cai22NCElectric}.  

\begin{figure}[!tp] 
	\centering
	\includegraphics[width=0.48\textwidth]{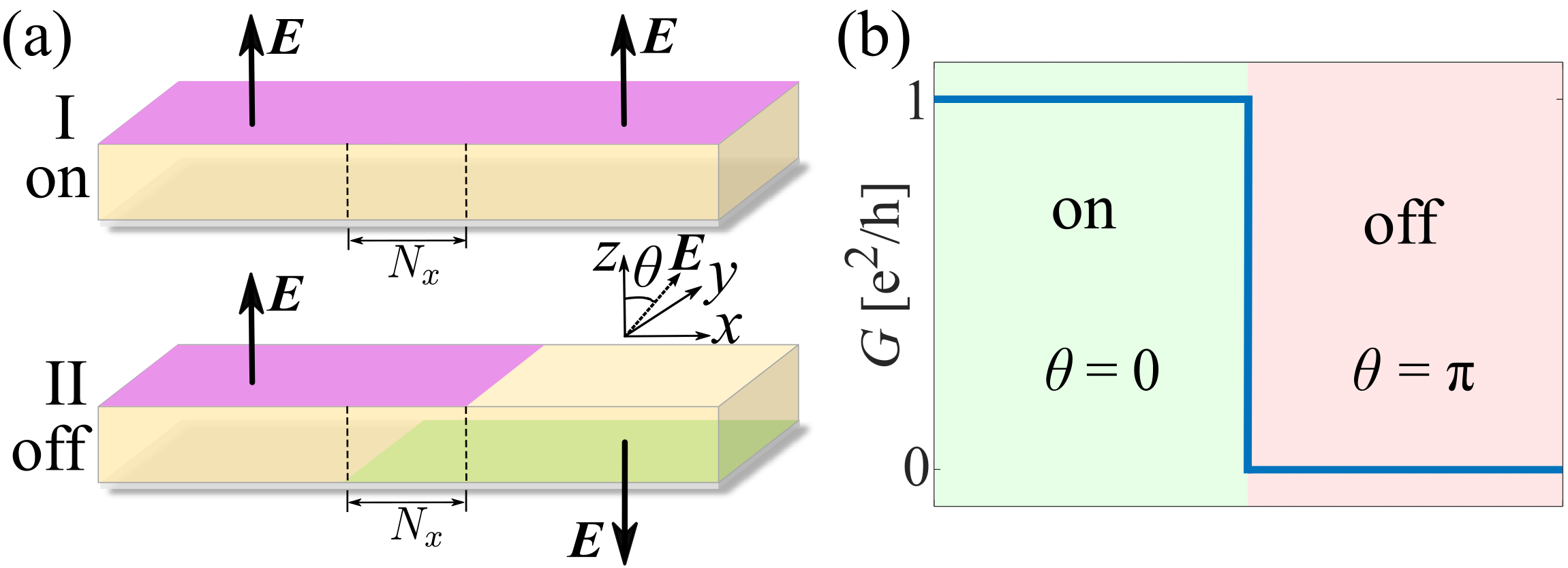}\\
	\caption{(a) Schematic of a magnetic topological transistor based on magnetic TIs. The magnetic TI (yellow) is grown on a substrate (grey). Two independent electric fields are applied to two neighboring regions. The middle region marked by two dashed lines is free of external electric fields. The external electric field in the left region is fixed along $z$-direction. The angle $\theta$ describes the relative angle of the two independent electric fields. For case I, the electric field in the right region is parallel to that in the left region and the transistor is in the ``on" state. For case II, the electric field in the right region is antiparallel to that in the left region and the transistor is in the ``off" state. The purple and green colors represent the dominant surface states crossing the Fermi level at top and bottom of the sample, respectively. (b) Two-terminal conductance of the magnetic topological transistor that switches from ``on" (quantized conductance) to ``off" (zero conductance) as the relative direction of the electric fields changes from parallel ($\theta=0$) to antiparallel ($\theta=\pi$). 
}\label{Fig:Normal_junction_setup}
\end{figure}

In this work, we propose a new on/off switching mechanism to realize a magnetic topological transistor based on the antiferromagnetic TI MnBi$_2$Te$_4$. 
This mechanism explores three crucial ingredients in MnBi$_2$Te$_4$: (i) topological Dirac surface states, (ii) intrinsic exchange fields, and (iii) layer degrees of freedom.
To elucidate the physical picture, we first construct an effective model for MnBi$_2$Te$_4$ thin films in presence of external electric fields. We then show that manipulating the direction of electric fields allows us to selectively address the transport of top and bottom Dirac surface states. Exploiting this layer degree of freedom, we propose a two-terminal magnetic topological transistor. The ``on" state (quantized conductance) and the ``off" state (zero conductance) of this device are selected by the relative directions of two adjacent electric fields [Figs.~\ref{Fig:Normal_junction_setup}(a)-(b)]. The physical reason is that we are able to guide the electron transport from the top surface state of the left region to either the top surface state of the right region (``on" state) or to the bottom surface state of the right region (``off" state).  We show below that high on-off ratios ($\sim 10^7$) of the magnetic topological transistor can be achieved when the antiferromagnetic MnBi$_2$Te$_4$ films satisfy one of two criteria: (i) The films are thick enough to avoid hybridization of top and bottom surface states. (ii) The films have compensated antiferromagnetic order. In the latter case, the transistor can tolerate a considerable hybridization of top and bottom surface states due to their opposite Berry curvature.
Our proposal requires electric control of MnBi$_2$Te$_4$, which has been demonstrated in recent experiments \cite{Gao21NLayer,Cai22NCElectric}.

\section{Effective model for surface states}
To demonstrate the control of layer degrees of freedom by electric fields, we first construct an effective model of antiferromagnetic MnBi$_2$Te$_4$ thin films in presence of an electric field. MnBi$_2$Te$_4$ can be viewed as a TI with intrinsic antiferromagnetic order due to an exchange field~\cite{Zhang19PRLTopological, Zhang20PRLMobius,Zhang09NPTopological}, which breaks time-reversal symmetry.  
In practice, the electric field can be applied by dual gate technology \cite{Gao21NLayer}. Without loss of generality, we assume that the electric field is applied along $z$ direction and can be described by an electric potential $V_E$, which is an odd function of $z$, i.e., $V_E(-z)=-V_E(z)$. This corresponds to symmetric gating at top and bottom surfaces. Asymmetric gating affects our results quantitatively but not qualitatively.

We start from the bulk Hamiltonian of  three-dimensional (3D) TIs and derive the four lowest-energy eigenstates at the $\Gamma$ point as a basis~\cite{Sun20PRBAnalytical, Lu13PRLQuantum,Shan10NJPEffective,Lu10PRBMassive}. Then, we project the antiferromagnetic order and the electric potential into this basis. The resulting effective model for the MnBi$_2$Te$_4$ thin films in presence of the electric field can be written as [see Appendix \ref{sec:SM_effective_model} for more details]
\begin{align}
	H = H_0 + H_{ex} + H_{V},
	\label{model}
\end{align}
where $H_0$ describes the Dirac surface states of TIs given by
\begin{align} \label{Eq:Effective_H0}
	H_0 = h({\bf k}) \tau_z \otimes \sigma_z - \gamma \tau_0 \otimes (k_x \sigma_y - k_y \sigma_x)
\end{align}
with $h({\bf k})={\Delta}/{2} - B k^2$ and $k^2 = k_{x}^2 + k_{y}^2$. 
$\tau_0$ and $\sigma_0$ are $2\times2$ identity matrices and $\tau_i$ and $\sigma_i$ with $i=x,y,z$ are Pauli matrices. The basis is \{$|+,\uparrow\rangle, |-,\downarrow\rangle, |-,\uparrow\rangle, |+,\downarrow\rangle$\} with $|\pm,\uparrow (\downarrow)\rangle=\frac{1}{\sqrt{2}}\left(|t,\uparrow (\downarrow)\rangle \pm |b,\uparrow (\downarrow)\rangle \right)$ where $t (b)$ represents top (bottom) surface states and $\uparrow (\downarrow)$ represents spin up (down) states. $\Delta,B$ and $\gamma$ are model parameters that depend on the thickness of the films [see Appendix \ref{sec:SM_effective_model}]. For thick films, both $\Delta$ and $B$ approach zero. 

The second term in Eq.~\eqref{model}, i.e., $H_{ex}$,  corresponds to the effective exchange field of the surface states. It opens a band gap in the spectrum of the Dirac surface states. Notably, the effective exchange field is different for even- and odd-layer thin films because of the antiferromagnetic order in the bulk. For even-layer films, the magnetization is compensated, whereas there is net magnetization for odd-layer films. As a result, $H_{ex}$  reads
\begin{align}
	H_{ex}^{\rm{odd}} = m_1 \tau_0 \otimes \sigma_z
\end{align}
for odd-layer films, and 
\begin{align}
	H_{ex}^{\rm{even}} = m_2 \tau_x \otimes \sigma_z 
\end{align}
for even-layer films, where $m_1$ and $m_2$ are the corresponding strengths of the effective exchange field. The last term $H_V$ in Eq.~\eqref{model} stems from the electric field. It reflects the structure inversion asymmetry (SIA) of the two surfaces described by 
\begin{align}
	H_{V}= V \tau_x \otimes \sigma_0\label{Eq:Vfield}
\end{align}
with $V$ being the SIA strength.

\begin{figure}[!tp]
	\centering
	\includegraphics[width=0.48\textwidth]{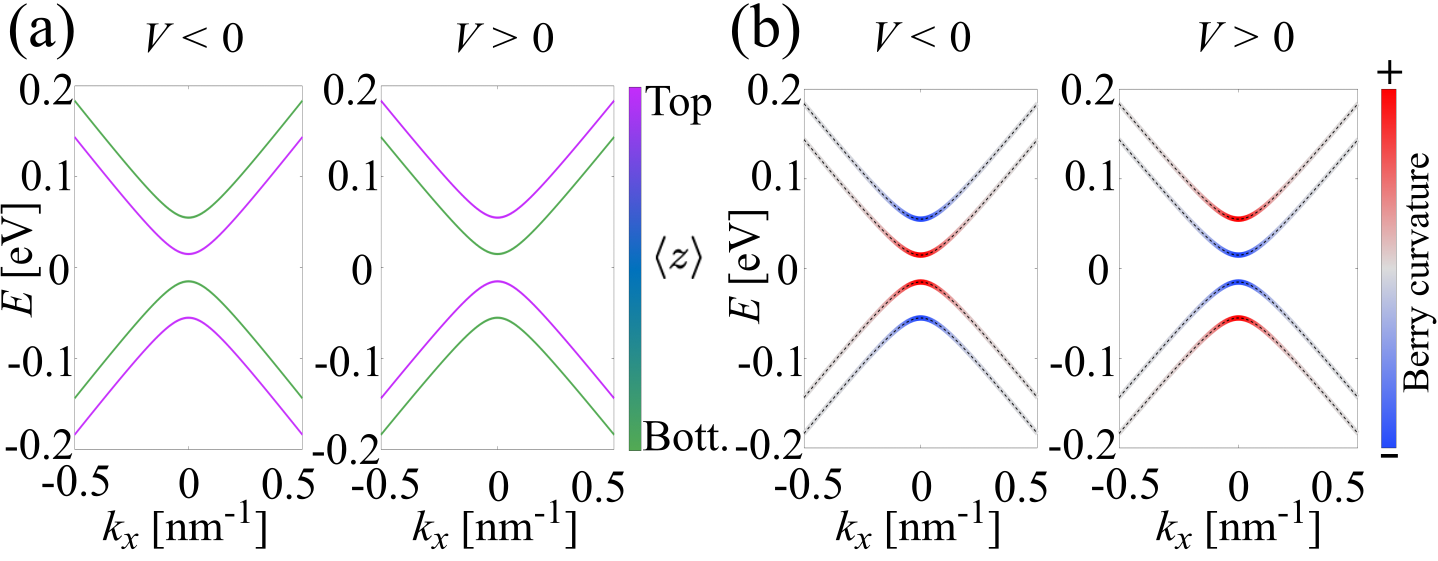}\\
	\caption{(a) Spectra for antiferromagnetic MnBi$_2$Te$_4$ films color-coded by the position expectation value $\langle z \rangle$ along $z$-direction in presence of opposite electric field directions. (b) Berry curvature distributions for even-layer films. Here we set $k_y=0$. The chosen parameters are adopted to the material MnBi$_2$Te$_4$ as described in Appendix \ref{sec:SM_effective_model}.}\label{Fig:AFM_spectrum}
\end{figure}

The SIA term $H_V$ induced by external electric fields can shift top and bottom surface states relative to each other in energy due to the potential difference at opposite surfaces, illustrated by the color-coded position expectation value $\langle z \rangle$ for the low-energy bands in Fig.~\ref{Fig:AFM_spectrum}(a). 
Reversing the direction of the electric field, this alters the shift pattern in an opposite way. Note that the shift patterns of the surface energy bands are the same in films with odd and even layers. This can be understood by recasting the effective model in the basis of top and bottom surface states [see Appendix \ref{sec:SM_effective_model} for more details], i.e.,
\begin{align}
	h_{s}= s V + m_j s^{j-1} \sigma_z + s\gamma (k_y \sigma_x - k_x \sigma_y), \label{h_cone}
\end{align}
where $s=\pm$ denote top and bottom surfaces, respectively, and $j=1(2)$ indicates odd(even)-layer films.
The new basis is \{$|t,\uparrow\rangle, |t,\downarrow\rangle, |b,\uparrow\rangle, |b,\downarrow\rangle$\}. 
Note that the spin-momentum locking term in Eq.~(\ref{h_cone}) is crucial for the robustness of surface state transport.  For simplicity, we assume in Eq.~(\ref{h_cone}) that the films are thick enough such that the hybridization of top and bottom surface states can be ignored (i.e., $\Delta=B=0$). 
Evidently, the SIA strength $V$ has opposite signs for top and bottom surface states. Hence, it shifts the Dirac bands in an opposite way. If the Fermi level is placed to cross the lowest conduction band, then flipping the direction of the external electric field (changing the sign of $V$), this selects topological surface states from opposite surfaces. 

Furthermore, the Berry curvature distributions of the surface states are strongly influenced by the SIA term $H_V$.
For even-layer films, in the absence of an electric field ($V=0$), the Berry curvature of the lowest-energy band is zero due to the presence of PT symmetry (i.e., combined space inversion and time-reversal symmetry)~\cite{Du20PRRBerry}. When the electric field is present ($V \neq 0$), PT symmetry is broken. Hence, the degeneracy of the energy bands is lifted and the Berry curvature of each band becomes finite. In contrast, for odd-layer films, the Berry curvature is always nonzero due to the breaking of PT symmetry~\cite{Li19SAIntrinsic} no matter whether the electric field is present or not. Moreover, the Berry curvature of the even-layer films is layer-locked for conduction (or valence) bands~\cite{Gao21NLayer}, illustrated in Figs.~\ref{Fig:AFM_spectrum}(a) and \ref{Fig:AFM_spectrum}(b). 

\section{Magnetic topological transistor}
We propose a magnetic topological transistor based on the unique properties described above [see Fig.~\ref{Fig:Normal_junction_setup}(a) for a schematic]. The two side regions of the junction under the influence of external electric fields are connected to source and drain of the transistor. The middle region is free of external fields. Considering a finite-size 3D slab geometry, the topological surface bands evolve to quasi-1D spectra along the longitudinal direction. The relative energy separation of top and bottom quasi-1D spectra can be controlled by electric fields. This feature selects the layer degrees of freedom. The transistor can switch between ``on'' and ``off'' states depending on the relative direction of the electric fields, as illustrated in Fig.~\ref{Fig:Normal_junction_setup}(b). When the electric fields are parallel, the transistor is in the ``on" state with quantized conductance $G \approx e^2/h$ considering a single quasi-1D spectrum at the Fermi level. In contrast, when the electric fields are antiparallel, the transistor is in the ``off" state with vanishing conductance $G \approx 0$. 

\begin{figure}[!tp]
	\centering
	\includegraphics[width=0.48\textwidth]{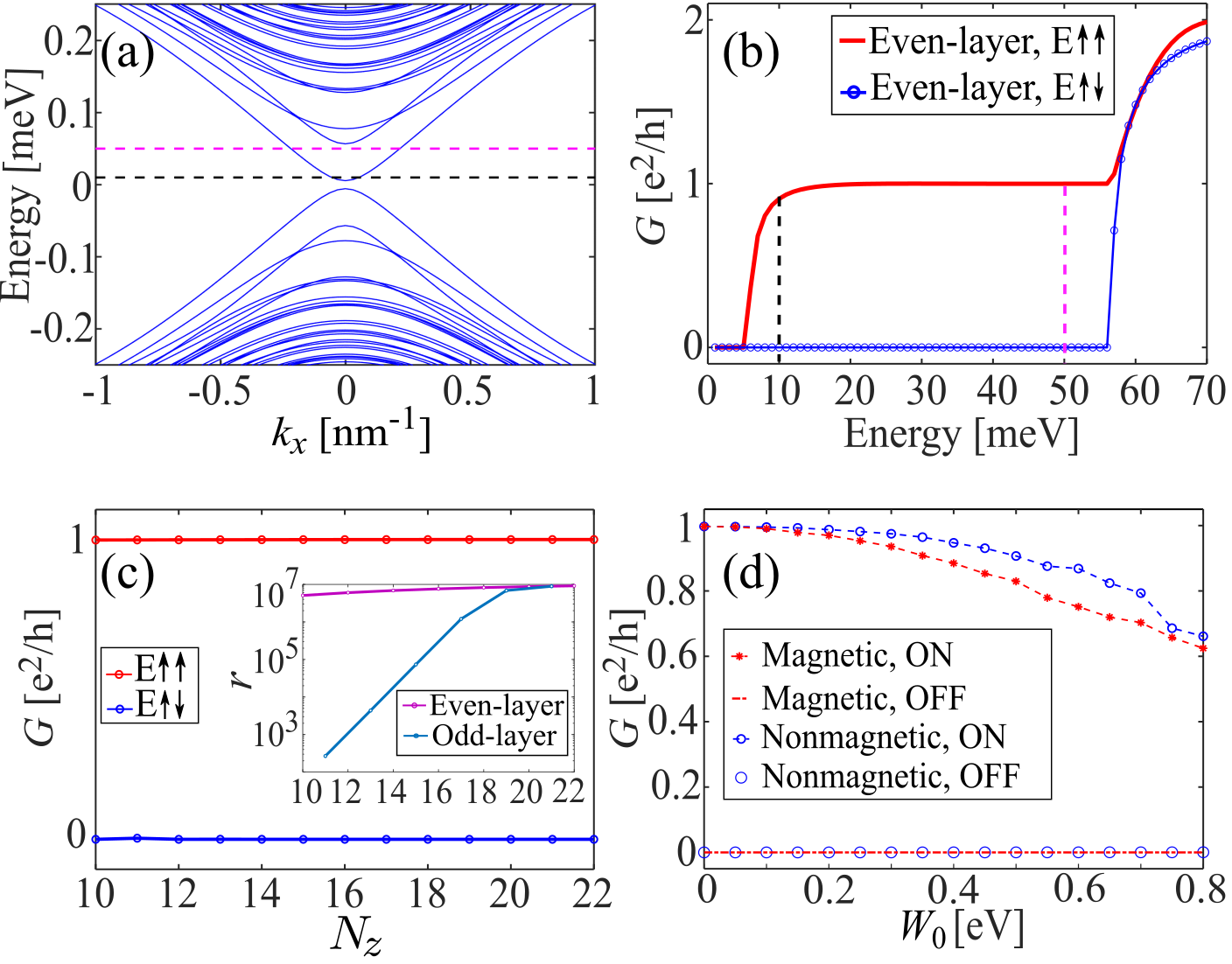}\\
	\caption{(a) Spectrum of even-layer 3D slab with $N_z=12$. The black and purple dashed lines represent Fermi energies at $E_F$=10, 50 meV, respectively. (b) Conductance $G$ of the magnetic topological transistor as a function of Fermi energy for different external field configurations with $N_z=12$ (even layers). 
	The symbols E$\uparrow \uparrow$ and E$\uparrow \downarrow$ represent the relative directions of the electric field for left and right leads to be parallel and antiparallel, respectively. The black and purple dashed lines represent Fermi energies at $E_F$=10, 50 meV, respectively. (c) Conductance $G$ as a function of junction thickness $N_z$ with $E_F$=30 meV. The inset shows the on-off ratio $r\equiv G_{\mathrm{on}}/G_{\mathrm{off}}$ as a function of $N_z$. (d) Conductance $G$ as a function of magnetic and nonmagnetic disorder strength $W_0$ with $N_z=10$. The spin polarization of the magnetic disorder is along $z$ direction. The Fermi level is fixed at $E_F$=25 meV. Open (periodic) boundary conditions are imposed in $z(y)$ direction.
	The other parameters in all plots are specified in \cite{Parameters_transistors}.
	}\label{Fig:AFM_Normal_conductance}
\end{figure}

The working principle of this transistor is based on layer degrees of freedom. At a given Fermi level, the top or bottom surface quasi-1D spectrum can solely be responsible for transport since the electric field separates different bands in energy [see Fig.~\ref{Fig:AFM_Normal_conductance}(a)].
We set the direction of electric field on the left-hand side along $+{\bm{z}}$ direction such that the Fermi level only crosses the spectrum of the top surface. For the middle region of the transistor, the Fermi level crosses the spectrum of both top and bottom surfaces. Eventually, the nature of the conducting surface spectrum at the Fermi level on the right-hand side matters. If the top surface state 
on the right-hand side is responsible for transport [case I in Fig.~\ref{Fig:Normal_junction_setup}(a)], then electrons traverse the junction easily and the conductance approaches  $G\approx e^2/h$. Otherwise [case II in Fig.~\ref{Fig:Normal_junction_setup}(a)], the conductance $G$ approaches zero if the thickness of the middle region is large enough such that hybridization of top and bottom surface states is negligible. Note that although the surface states are gapped by the magnetic order, they inherit spin-momentum locking described by Eq.~(\ref{Eq:Effective_H0}). Therefore, the back scattering from impurities is substantially suppressed.

\section{Numerical simulations}
Up to now, we have analyzed the mechanism for our proposed magnetic topological transistor from a thick-film perspective. To directly analyze the performance of the transistor for any film thickness, we perform numerical calculations based on the discretized 3D bulk Hamiltonian on a cubic lattice [see Appendix \ref{sec:SM_lattice_model}], by employing the Landauer formalism \cite{landauer1970electrical,buttiker1986four,Datta97Electronic}. We choose typical bulk parameters that are adopted to the material MnBi$_2$Te$_4$ based on {\it{ab initio}} calculations \cite{Zhang19PRLTopological}.  The chosen sample geometry consists of a cuboidal central region and two semi-infinite leads as source and drain. Exploiting the recursive Green function technique \cite{MacKinnon85ZPB-CMCalculation}, the conductance of the setup can be evaluated by 
\begin{align}
	G= \frac{e^2}{h} \mathrm{Tr}[\Gamma_{L} G^r\Gamma_R G^a],
\end{align}
where $\Gamma_{L/R}=i(\Sigma_{L/R}^r-\Sigma_{L/R}^a)$ are the linewidth functions with $\Sigma_{L/R}$ the self-energy due to the coupling to the left/right lead. $G^r(G^a)$ is the retarded (advanced) Green function of the central region.

Figures~\ref{Fig:AFM_Normal_conductance}(a) and~\ref{Fig:AFM_Normal_conductance}(b) illustrate that the magnetic topological transistor works perfectly in a wide energy window, in which the conductance $G$ approaches $e^2/h$ for the ``on" state (parallel configuration E$\uparrow\uparrow$) and decreases to nearly zero for the ``off" state (antiparallel configuration E$\uparrow\downarrow$). Importantly, the range of this effective energy window is controllable by the strength of electric fields. This result is consistent with the phenomenological expectations discussed before. The on-off ratio $r\equiv G_{\mathrm{on}}/G_{\mathrm{off}}$ of this transistor can reach $10^7$, as shown in the inset of Fig.~\ref{Fig:AFM_Normal_conductance}(c). When the Fermi level lies at the bottom of the quasi-1D spectrum, the value of $r$ decreases due to diminishing spin-momentum locking. 

To obtain a large on-off ratio $r$, the thickness of the junction $N_z$ should be large enough (i.e, $N_z\ge10$) to avoid hybridization of top and bottom surface states, as shown in the inset of Fig.~\ref{Fig:AFM_Normal_conductance}(c). The on-off ratio $r$ increases as $N_z$ grows. Odd-layer devices only perform well for thick films while even-layer devices are less sensitive to the film thickness. 
Moreover, to illustrate the robustness of the magnetic topological transistor against perturbations, Anderson-type disorder is introduced through random on-site potential with a uniform distribution within $[-W_0/2, W_0/2]$, where $W_0$ denotes the disorder strength. It is evident from Fig.~\ref{Fig:AFM_Normal_conductance}(d) that the proposed magnetic topological transistor persists to perform well in presence of disorders. The strong protection against perturbations comes from spin-momentum locking inherited from topological surface states. Moreover, the conductance is also robust against weak magnetic disorder. 
Note that the magnetic disorder in antiferromagnetic MnBi$_2$Te$_4$ may be formed by Mn sites filled with Bi atoms or Bi sites occupied by Mn atoms~\cite{Zeugner19CMChemical,Shikin21PRBSampledependent,Garnica22nQMNative}. Therefore, the spin polarization of the magnetic disorder is considered along $z$-direction in the calculation. For simplicity, we discuss the functionality of the magnetic topological transistor with a single conducting channel. When more channels are involved, the transistor works even better with higher on-off ratios [see Appendix \ref{sec:SM_transport}]. In our simulations, all the parameters are based on the \textit{ab initio} calculations \cite{Zhang19PRLTopological}, which should apply to MnBi$_2$Te$_4$.  

\begin{figure}[!tp]
	\centering
	\includegraphics[width=0.48\textwidth]{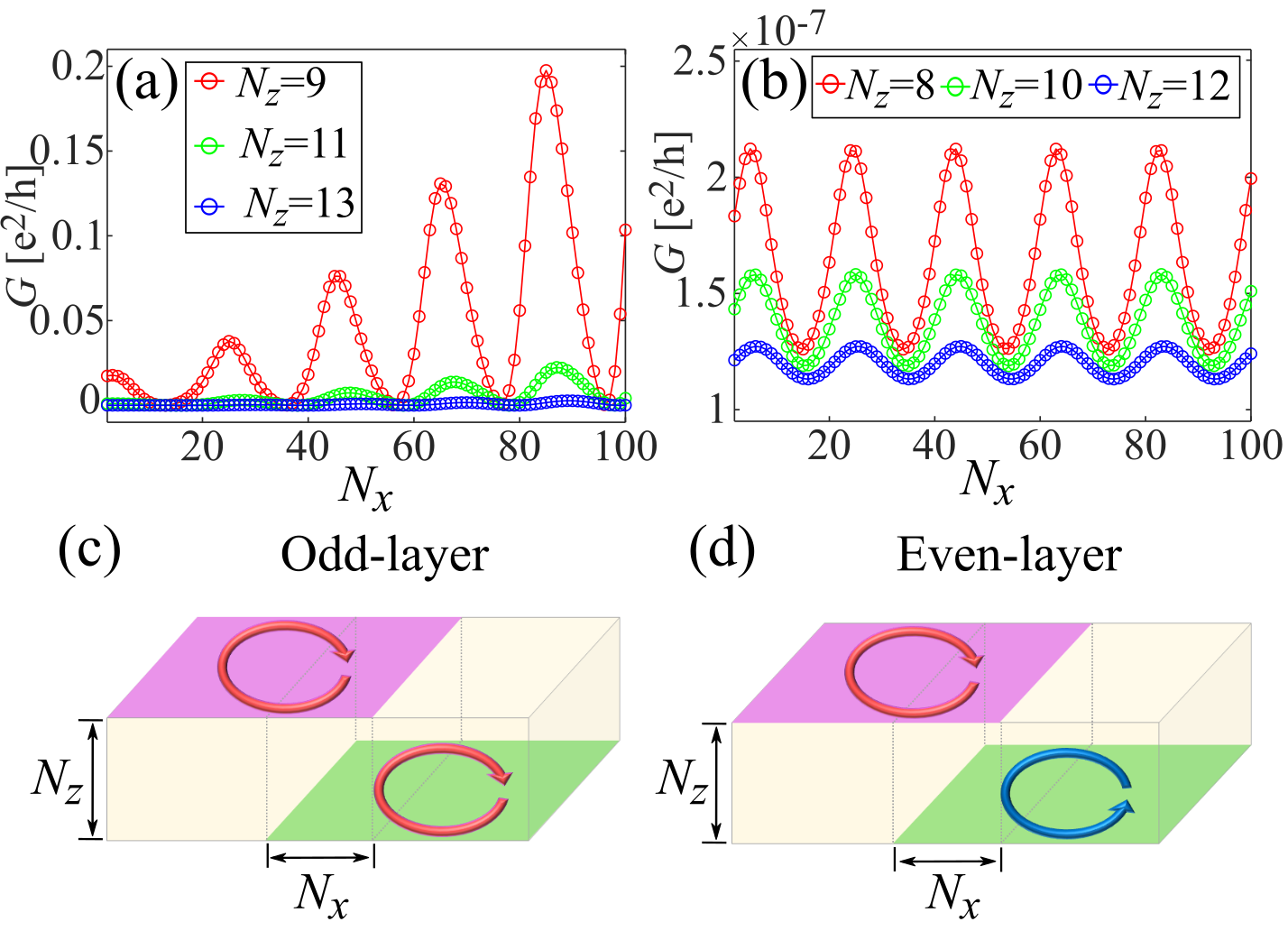}\\
	\caption{(a) Conductance $G$ as a function of $N_x$ for case II in Fig.~\ref{Fig:Normal_junction_setup}(a) with $N_z=9$, 11, and 13, respectively. (b) Conductance $G$ as a function of $N_x$ for case II in Fig.~\ref{Fig:Normal_junction_setup}(a) with $N_z=8$, 10, and 12, respectively. We choose Fermi energy $E_F$=30 meV, and other parameters are the same as in Fig.~\ref{Fig:AFM_Normal_conductance}. (c) Illustration of the Berry curvature for top surface (purple region) and bottom surface (green region) for antiparallel configuration E$\uparrow\downarrow$ for odd-layer thin films. Red arrows and blue arrows represent opposite Berry curvature. $N_x$ is the length of the middle region and $N_z$ is the thickness of the thin films. (d) Illustration of the Berry curvature for top surface and bottom surface for antiparallel configuration E$\uparrow\downarrow$ for even-layer thin films.}\label{Fig:Conductance_Nx_var}
\end{figure}

\section{Berry curvature features} 
For thick films, the hybridization of top and bottom surface states is negligible. However, as the film thickness decreases, the hybridization of top and bottom surfaces becomes pronounced. The tunneling of the surface electrons from top to bottom diminishes the functionality of the magnetic topological transistor. Figure~\ref{Fig:Conductance_Nx_var}(a) shows that the conductance $G$ oscillates (as a function of $N_x$) with finite amplitude as the thickness of the transistor $N_z$ decreases in odd-layer thin films. In striking contrast, the conductance $G$ of the even-layer thin films stays close to zero (of the order of 10$^{-7}$ e$^2$/h) even when $N_z$ is quite small [Fig.~\ref{Fig:Conductance_Nx_var}(b)]. 
The different behaviors between odd and even layer cases originate from the spin polarization of top and bottom surface states. For a given Fermi energy intersecting the conduction band, the spin polarization is the same for odd-layer films but opposite for even-layer case in agreement with the direction of the magnetization. This is featured by the same Berry curvature distribution for odd-layer system [Fig.~\ref{Fig:Conductance_Nx_var}(c)] and the opposite Berry curvature distribution for even-layer system [Fig.~\ref{Fig:Conductance_Nx_var}(d)]. We show in Appendix \ref{sec:SM_transport} that this difference of the Berry curvature distribution crucially affects the transport properties in the ``off" state. Consequently, even-layer thin film devices are superior as compared to odd-layer thin film devices.

\section{Conclusion}
We have developed an effective model for the antiferromagnetic TI MnBi$_2$Te$_4$ in presence of electric fields to demonstrate that electric fields can be exploited to selectively guide surface state transport. We have proposed a magnetic topological transistor based on three crucial ingredients: topology (for the emergence of surface states with strong spin-momentum locking), exchange fields (to gap the surface states), and electric-field-controllable layer degrees of freedom (to select top and bottom surface states). The proposed magnetic topological transistor is robust against disorder due to strong spin-momentum locking of the surface states. It shows large on-off ratios when the magnetic TI films are thick enough to avoid hybridization or have fully compensated magnetic order. Our work suggests a new switching mechanism for transistors exploiting layer-selective transport.

\begin{acknowledgments}
We thank Rui Chen for valuable discussions. This work was supported by the DFG (SPP1666 and SFB1170 “ToCoTronics”), the W\"urzburg-Dresden Cluster of Excellence ct.qmat, EXC2147, Project No. 390858490, and the Elitenetzwerk Bayern Graduate School on “Topological Insulators”. We thank the Bavarian Ministry of Economic Affairs, Regional Development and Energy for ﬁnancial support within the High-Tech Agenda Project “Bausteine f\"ur das Quanten Computing auf Basis topologischer Materialen”. S.B.Z acknowledges the
support by the UZH Postdoc Grant. H.Z.L. acknowledges support by the National Natural Science Foundation of China (Grant No. 11925402).
\end{acknowledgments}

\appendix

\section{Effective model for surface states} \label{sec:SM_effective_model}

\subsection{3D bulk Hamiltonian for antiferromagnetic MnBi$_2$Te$_4$ in presence of an electric field}
The 3D bulk Hamiltonian in presence of an external electric field for antiferromagnetic topological insulators (TIs) MnBi$_2$Te$_4$ reads 
\begin{eqnarray} 
	\mathcal{H}(\boldsymbol{k}) = \mathcal{H}_N(\boldsymbol{k}) + \mathcal{H}_X(z) + V_E(z), \label{Eq_3D_Hamiltonian}
\end{eqnarray}
where $\mathcal{H}_N(\boldsymbol{k})$ is the nonmagnetic part, $\mathcal{H}_X(z)$ is the exchange field that accounts for the out-of-plane antiferromagnetic order, and $V_E(z)$ is the electric potential induced by the external electric field. The non-magnetic part is given by \cite{Zhang19PRLTopological}
\begin{equation} \label{Eq:SM_bulk_Hamiltonian}
	\mathcal{H}_N (\boldsymbol{k}) = \epsilon_{0}(\boldsymbol{k}) +
	\begin{bmatrix}
		M(\boldsymbol{k}) & A_{1} k_{z}        & 0                 & A_{2} k_{-} \\
		A_{1} k_{z}       & -M(\boldsymbol{k}) & A_{2} k_{-}       & 0 \\
		0                 & A_{2} k_{+}        & M(\boldsymbol{k}) & -A_{1} k_{z} \\
		A_{2} k_{+}       & 0                  & -A_{1} k_{z}      & -M(\boldsymbol{k})
	\end{bmatrix} 
\end{equation}
with $k_{\pm}=k_{x} \pm i k_{y}$, $\epsilon_{0}(\boldsymbol{k})=  C_{0} + D_{1}k_{z}^{2} + D_{2}(k_{x}^{2}+k_{y}^{2})$ and $M(\boldsymbol{k})=M_{0} - B_{1}k_{z}^{2} - B_{2}(k_{x}^{2}+k_{y}^{2}) $. $C_0$, $D_i$, $M_0$, $B_i$ and $A_i$ are model parameters with $i=1,2$. The basis of the Hamiltonian is \{$ |{P1_{z}^{+}, \uparrow}\rangle, |{P2_{z}^{-}, \uparrow}\rangle,|{P1_{z}^{+}, \downarrow}\rangle,|{P2_{z}^{-}, \downarrow}\rangle $\}.

The exchange field reads
\begin{eqnarray}
	\mathcal{H}_X(z)
	= m_z(z) \sigma_z \otimes \tau_0,
\end{eqnarray}
where $m_z(z) = -m_0 \, {\rm{sin}} (\pi z/ d )$ is the magnetization energy along $z$-direction with $m_0$ the amplitude of the intralayer ferromagnetic order. $d$ is the thickness of a septuple layer. $\sigma_{z}$ is the Pauli $z$ matrix for spin degrees of freedom, and $\tau_0$ is a 2$\times$2 identity matrix for orbital degrees of freedom. 

We assume the electric potential $V_E(z)$ is an odd function of $z$, namely, $V_E(-z)=-V_E(z)$, which corresponds to symmetric gating. The qualitative findings of ours are not affected by this choice.

\subsection{Wave functions and eigen energies for MnBi$_2$Te$_4$ thin films}
Consider a MnBi$_2$Te$_4$ thin film with thickness $L$ along $z$ direction and label bottom and top boundaries as $-L/2$ and $L/2$, respectively. First, we derive the surface eigenstates of the nonmagnetic part at the $\Gamma$ point ($k_x=k_y=0$), which form the basis for the effective model. Replacing $k_z$ by $-i\partial_z$, since $k_z$ is no longer a good quantum number, and taking $k_x=k_y=0$ in $\mathcal{H}_N(\boldsymbol{k})$, we arrive at $\mathcal{H}_N(-i\partial_z)$, where
\begin{eqnarray} \label{Eq:SM_Hamiltonian_partial_z}
	\mathcal{H}_N(-i\partial_z)
	=&
	\begin{bmatrix}
		h_N(A_1) & 0 \\
		0      & h_N(-A_1)
	\end{bmatrix}
\end{eqnarray}
with $h_N(A_1)=C-D_1 \partial_z^2 + (M+B_1 \partial_z^2)\sigma_z - iA_1 \partial_z \sigma_x$.
The general eigenstate for $h_N(A_1)$ reads 
\begin{eqnarray}
	\psi (z)
	=&
	C_{\alpha,\beta} \psi_{\alpha, \beta} e^{\beta \lambda_{\alpha} z},
\end{eqnarray}
where 
\begin{eqnarray}
	\psi_{\alpha, \beta}
	=
	\begin{bmatrix}
		-D_{+}\lambda_{\alpha}^2 + l_{-} - E \\
		iA_{1} \beta \lambda_{\alpha}
	\end{bmatrix}, \\
	\lambda_{\alpha}=\sqrt{\frac{-F+(-1)^{\alpha -1}\sqrt{R}}{2(D_1^2 - B_1^2)}} 
\end{eqnarray}
with $\alpha=1,2$, $\beta=\pm$, $D_{+}=D_{1} + B_{1}$, $l_{-}=C_0 - M_0$, $F=A_1^2-2 B_1 M_0 + 2 D_1 (E_0-C_0)$, and $R=F^2 - 4 \left(D_1^2-B_1^2\right) (E-C_0-M_0) (E-C_0+M_0)$.

Using open boundary conditions at $z=\pm L/2$, namely, $\psi(\pm L/2)=0$, 
we obtain the transcendental equations for the eigenenergies
\begin{widetext}
	\begin{eqnarray}
		E_{+} =& \dfrac{  (-D_{+}\lambda_{1}^2 + l_{-})  \lambda_{2} {\rm tanh}(\lambda_1 L / 2) - (-D_{+}\lambda_{2}^2 + l_{-})  \lambda_{1} {\rm tanh}(\lambda_2 L / 2)  }{ \lambda_{2} \, {\rm tanh}(\lambda_1 L / 2) - \lambda_{1} \, {\rm tanh}(\lambda_2 L / 2)  },  \\
		E_{-}  =&  \dfrac{  (-D_{+}\lambda_{1}^2 + l_{-})  \lambda_{2} {\rm tanh}(\lambda_2 L / 2) - (-D_{+}\lambda_{2}^2 + l_{-})  \lambda_{1} {\rm tanh}(\lambda_1 L / 2)  }{ \lambda_{2} \, {\rm tanh}(\lambda_2 L / 2) - \lambda_{1} \, {\rm tanh}(\lambda_1 L / 2)  }.
	\end{eqnarray}
\end{widetext}
Moreover, we find the wave functions from the general solution as 
\begin{eqnarray}
	\varphi(A_1)
	=&
	C_+
	\begin{bmatrix}
		-D_{+} f_{-}^{+}(z) \eta^{+}  \\
		iA_{1} f_{+}^{+}(z)
	\end{bmatrix}, \\
	\chi(A_1)
	=&
	C_{-}
	\begin{bmatrix}
		-D_{+} f_{+}^{-}(z)\eta^{-} \\
		iA_{1} f_{-}^{-}(z)
	\end{bmatrix}
\end{eqnarray}
with
\begin{eqnarray}
	f_{+}^{\pm}(z) =
	\frac{{\rm{cosh}}(\lambda_1 z)}{{\rm cosh}(\lambda_1 \frac{L}{2})}
	- \frac{ {\rm{cosh}}(\lambda_2 z) }{ {\rm cosh}(\lambda_2 \frac{L}{2})} \Biggl\lvert_{E=E_{\pm}}, \\
	f_{-}^{\pm}(z)
	=
	\frac{{\rm{sinh}}(\lambda_1 z)}{ {\rm sinh}(\lambda_1 \frac{L}{2}) } -
	\frac{{\rm{sinh}}(\lambda_2 z)}{ {\rm sinh}(\lambda_2 \frac{L}{2}) } \Biggl\lvert_{E=E_{\pm}},  \\
	\eta^{+}
	=
	\frac{ \lambda_{2}^2 - \lambda_{1}^2 }{\lambda_{2} {\rm coth}(\lambda_2 \frac{L}{2}) - \lambda_{1} {\rm coth}(\lambda_1 \frac{L}{2})} \Biggl\lvert_{E=E_{+}}, \\
	\eta^{-}
	=
	\frac{\lambda_2^2 - \lambda_1^2}{\lambda_{2} {\rm tanh}(\lambda_2 \frac{L}{2}) - \lambda_{1} {\rm tanh}(\lambda_1 \frac{L}{2})} \Biggl\lvert_{E=E_{-}}. 
\end{eqnarray}
$C_+$ and $C_-$ are the normalization coefficients.
Thus, the wave functions for $\mathcal{H}_N(-i\partial_z)$ in Eq.(\ref{Eq:SM_Hamiltonian_partial_z}) read
\begin{eqnarray}
	\Phi_1 &=&
	\begin{bmatrix}
		\varphi(A_1) \\
		0
	\end{bmatrix},\,
	\Phi_2 =
	\begin{bmatrix}
		\chi(A_1) \\
		0
	\end{bmatrix},\\
	\Phi_3 &=&
	\begin{bmatrix}
		0 \\
		\varphi(-A_1)
	\end{bmatrix},\,
	\Phi_4 =
	\begin{bmatrix}
		0 \\
		\chi(-A_1)
	\end{bmatrix}.
\end{eqnarray}

\subsection{Effective model for MnBi$_2$Te$_4$ thin films in presence of an electric field}
Projecting the 3D bulk Hamiltonian for MnBi$_2$Te$_4$ to the basis \{$\Phi_1, \Phi_4, \Phi_2, \Phi_3$\}, we obtain the effective model as
\begin{eqnarray}
	H 
	&=&
	\int_{-\frac{L}{2}}^{\frac{L}{2}} dz
	[\Phi_1, \Phi_4, \Phi_2, \Phi_3]^{\dagger} \mathcal{H} (\boldsymbol{k}) [\Phi_1, \Phi_4, \Phi_2, \Phi_3] \nonumber \\
	&=& 
	H_0 + H_{ex} + H_{V},
\end{eqnarray}
where
\begin{align}
	H_0 = h(k) \tau_z \otimes \sigma_z - \gamma \tau_0 \otimes (k_x \sigma_y - k_y \sigma_x),
\end{align}
with  $h(k)={\Delta}/{2} - B k^2$, $\Delta = E_+ - E_-$, $k^2 = k_{x}^2 + k_{y}^2$, $ B = (\widetilde{B}_1 - \widetilde{B}_2)/2$, $ D = (\widetilde{B}_1 + \widetilde{B}_2)/2 - D_2$, 
$ 
\widetilde{B}_1
=
B_2 
\langle \varphi(A_1)|\sigma_z|\varphi(A_1)\rangle$, 
$
\widetilde{B}_2
=
B_{2} \langle \chi(A_1)|\sigma_z|\chi(A_1)\rangle$, $\gamma=-i\widetilde{A}_2$, and
$
\widetilde{A}_2
=
A_{2}\langle \varphi(A_1)| \sigma_x |\chi(-A_1)\rangle 
$.
For simplicity, we have ignored the energy shift $E_0$ and the particle-hole asymmetry term $Dk^2$, which implies that $\epsilon_{0}(\boldsymbol{k})=0$ in Eq.(\ref{Eq:SM_bulk_Hamiltonian}). 

The effective exchange field $H_{ex}(z)$ for odd-layer thin films reads
\begin{align}
	H_{ex}^{\rm{odd}}
	&=
	\int_{-\frac{L}{2}}^{\frac{L}{2}} dz
	[\Phi_1, \Phi_4, \Phi_2, \Phi_3]^{\dagger} \mathcal{H}_X(z) [\Phi_1, \Phi_4, \Phi_2, \Phi_3] \nonumber \\
	&= m_1 \tau_0 \otimes \sigma_z, \label{Eq_m1}
\end{align}
where
$m_1=
(-1)^{\frac{N_z-1}{2}}m_0 \langle \varphi(A_1)|{\rm{cos}} (\frac{\pi}{d} z)\sigma_0|\varphi(A_1)\rangle$ with $N_z$ the number of layers. 

The effective exchange field $H_{ex}(z)$ for even-layer thin films reads
\begin{align}
	H_{ex}^{\rm{even}}
	&=
	\int_{-\frac{L}{2}}^{\frac{L}{2}} dz
	[\Phi_1, \Phi_4, \Phi_2, \Phi_3]^{\dagger} \mathcal{H}_X(z) [\Phi_1, \Phi_4, \Phi_2, \Phi_3] \nonumber \\
	&= m_2 \tau_x \otimes \sigma_z, \label{Eq_m2}
\end{align}
where 
$m_2=
(-1)^{\frac{N_z}{2}-1} m_0 \langle \varphi(A_1)|{\rm{sin}}(\frac{\pi}{d} z)\sigma_0|\chi(A_1)\rangle$ with $N_z$ the number of layers. In Eqs.(\ref{Eq_m1}) and (\ref{Eq_m2}), $m_1$ and $m_2$ are the strengths of the effective exchange field for the surface states of odd- and even-layer thin films, respectively. $H_{ex}^{\rm{odd}}$ and $H_{ex}^{\rm{even}}$ are not the same because of the different profiles of bulk magnetization.

Structure inversion asymmetry (SIA) induced by the electric field is described by 
\begin{align}
	H_{V}
	&=
	\int_{-\frac{L}{2}}^{\frac{L}{2}} dz
	[\Phi_1, \Phi_4, \Phi_2, \Phi_3]^{\dagger} V_E(z) [\Phi_1, \Phi_4, \Phi_2, \Phi_3]  \nonumber \\
	&= V \tau_x \otimes \sigma_0,
\end{align}
where 
$
V
=
\langle \varphi(A_1)|V_E(z) \sigma_0 |\chi(A_1)\rangle 
$
is the strength of SIA.

In summary, the effective model for odd-layer MnBi$_2$Te$_4$ thin films in presence of electric fields reads
\begin{eqnarray}
	H_{\rm{odd}} =
	\begin{bmatrix}
		h(k)+m_1 & i \gamma k_{-} & V & 0 \\
		-i \gamma k_{+} & -h(k)-m_1 & 0 & V \\
		V & 0 & -h(k)+m_1 & i \gamma k_{-} \\
		0 & V & -i \gamma k_{+} & h(k)-m_1
	\end{bmatrix}, \nonumber 
\end{eqnarray}
with $h(k)={\Delta}/{2} - B k^2$, and $k^2 = k_{x}^2 + k_{y}^2$.

Likewise, the effective model for even-layer MnBi$_2$Te$_4$ thin films reads
\begin{eqnarray}
	H_{\rm{even}} =
	\begin{bmatrix}
		h(k) & i \gamma k_{-} & V+m_2 & 0 \\
		-i \gamma k_{+} & -h(k) & 0 & V-m_2 \\
		V+m_2 & 0 & -h(k) & i \gamma k_{-} \\
		0 & V-m_2 & -i \gamma k_{+} & h(k)
	\end{bmatrix}. \nonumber
\end{eqnarray}
Note that the basis of the model is a mixture of top and bottom surface states as \{$|+,\uparrow\rangle, |-,\downarrow\rangle, |-,\uparrow\rangle, |+,\downarrow\rangle$\} with $|\pm,\uparrow (\downarrow)\rangle=\frac{1}{\sqrt{2}}|t,\uparrow (\downarrow)\rangle \pm |b,\uparrow (\downarrow)\rangle$. 

The parameters of the effective model for antiferromagnetic MnBi$_2$Te$_4$ thin films used in Fig.~2 are shown below. Note that the parameters for the effective model are obtained from the bulk parameters extracted from {\it{ab initio}} calculations \cite{Zhang19PRLTopological}. For the even-layer film with $N_z=10$, the parameters are $\Delta=$$-0.34$ meV, $B=$1.31 meV$\cdot$nm$^2$, $\gamma=$319.64 meV$\cdot$nm,  $m_{1}=$0 meV, $m_2=$35.02 meV, $V=$20 meV. For the odd-layer film with $N_z=11$, the parameters are $\Delta=$$-0.16$ meV, $B=$0.67 meV$\cdot$nm$^2$, $\gamma=$319.64 meV$\cdot$nm,  $m_{1}=$35.02 meV, $m_2=$0 meV, $V=$20 meV.

\subsection{Effective model for MnBi$_2$Te$_4$ thick films}
For MnBi$_2$Te$_4$ thick films, the parameters $\Delta$ and $B$ approach zero, which implies that $h(k)=0$.
The unitary transformation is used to transform the basis as 
	\begin{align}
		\mathcal{\hat{H}}
		& =\sum \psi_1^{\dagger} H \psi_1   
		=\sum (U \psi_1)^{\dagger} U H U^{-1} (U \psi_1)  
	\end{align}
	with $U^{\dagger}U=1$. Thus, the new basis is $\psi_2=U \psi_1$. 
	Here, the unitary operator reads
	\begin{eqnarray}
	U &=& \frac{1}{\sqrt{2}}
	\begin{bmatrix}
		1 & 0 & 1 & 0 \\
		0 & 1 & 0 & 1 \\
		1 & 0 & -1 & 0 \\
		0 & -1 & 0 & 1
	\end{bmatrix}.
	\end{eqnarray}
	Thus the new basis is \{$|t,\uparrow\rangle, |t,\downarrow\rangle, |b,\uparrow\rangle, |b,\downarrow\rangle$\}.
The effective model for odd-layer MnBi$_2$Te$_4$ thick films in the new basis reads
\begin{eqnarray}\label{Eq:tpodd}
	UH_{\rm{odd}}U^{\dagger} &=&
	\begin{bmatrix}
		V+m_1 & i \gamma k_{-} & h(k) & 0 \\
		-i \gamma k_{+} & V-m_1 & 0 & h(k) \\
		h(k) & 0 & -V+m_1 & -i \gamma k_{-} \\
		0 & h(k) & i \gamma k_{+} & -V-m_1
	\end{bmatrix}. \nonumber
\end{eqnarray}

Similarly, the effective model for even-layer MnBi$_2$Te$_4$ thick films in the new basis reads
\begin{eqnarray}\label{Eq:tpeven}
	UH_{\rm{even}}U^{\dagger} &=&
	\begin{bmatrix}
		V+m_2 & i \gamma k_{-} & h(k) & 0 \\
		-i \gamma k_{+} & V-m_2 & 0 & h(k) \\
		h(k) & 0 & -V-m_2 & -i \gamma k_{-} \\
		0 & h(k) & i \gamma k_{+} & -V+m_2
	\end{bmatrix}. \nonumber
\end{eqnarray}

\begin{figure}[!tp]
	\centering
	\includegraphics[width=0.45\textwidth]{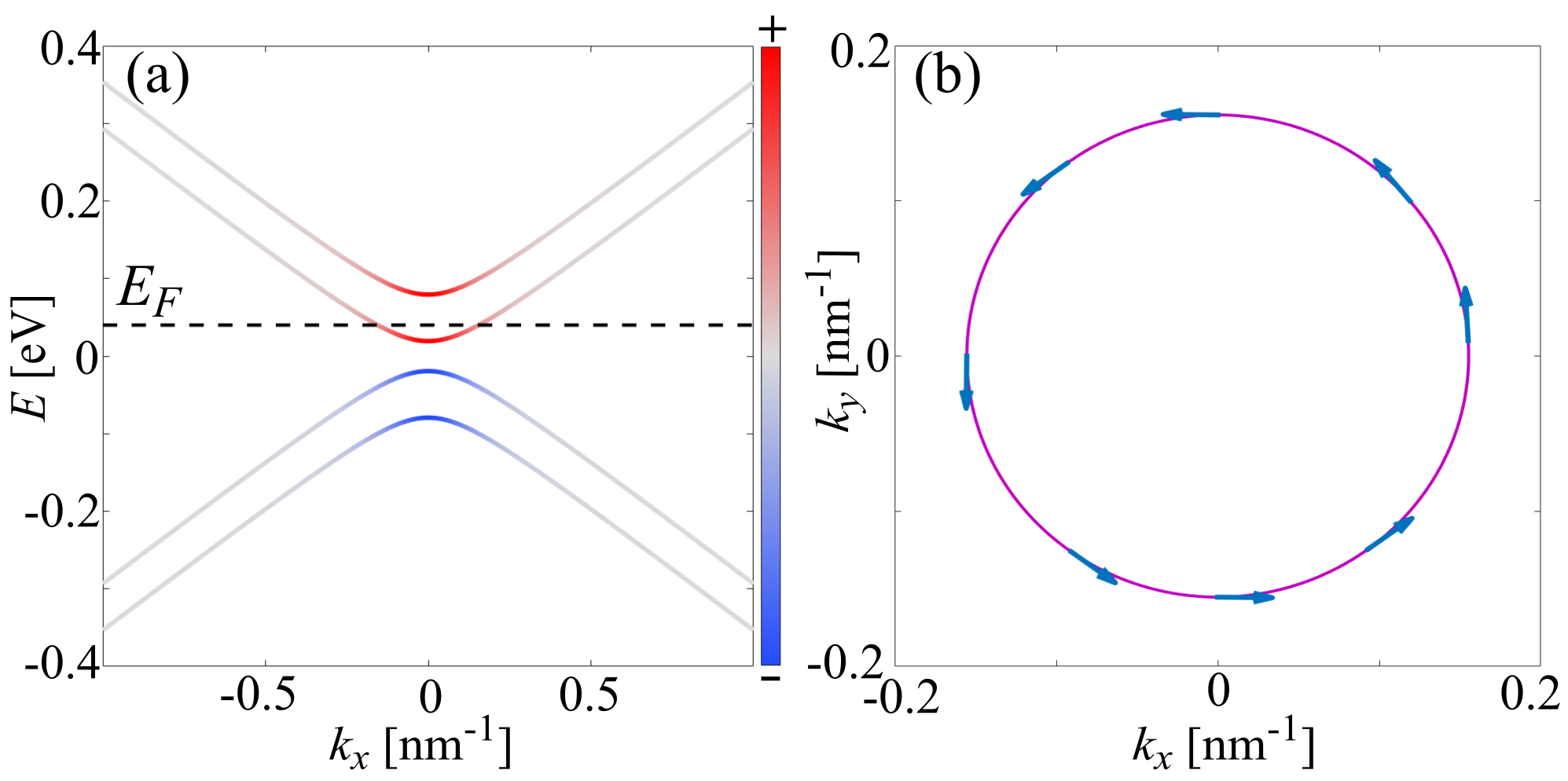}\\
	\caption{(a) Spectrum and Berry curvature distribution for odd-layer film with $N_z$=9 and $V$=30 meV. The dashed line indicates the Fermi energy $E_F$. (b) The spin texture of the energy cut in (a) with $E_F$=40 meV. } \label{Fig:effective_model}
\end{figure}

\subsection{Spin texture of the effective model}	
The spectrum of the four-band effective model for odd-layer thin films color-coded by the Berry curvature is shown in Fig.~\ref{Fig:effective_model}(a). We can see the Rashba-type spin texture in $k_x$-$k_y$ plane in Fig.~\ref{Fig:effective_model}(b), inherited from gapless topological surface states.

\section{Lattice model for magnetic TIs in presence of electric fields} \label{sec:SM_lattice_model}
The effective Hamiltonian of Eq.(\ref{Eq_3D_Hamiltonian}) can be regularized
on a cubic lattice by the substitutions $k_{j}\rightarrow\frac{1}{a_0}\sin(k_{j}a_0)$,
$k_{j}^{2}\rightarrow\frac{2}{a_0^{2}}[1-\cos(k_{j}a_0)]$, where $a_0$ is
the lattice constant and $j=x,y,z$. For simplicity, we take $a_0=1$ nm. The three lattice translation vectors are defined as $a_{1}=(1,0,0)^{T},a_{2}=(0,1,0)^{T},a_{3}=(0,0,1)^{T}$. The lattice model for antiferromagnetic MnBi$_2$Te$_4$ reads
\begin{eqnarray}
	H(\boldsymbol{k})=H_{0}(\boldsymbol{k})+H_{ex}+V_E,
\end{eqnarray}
where 
\begin{eqnarray}
	H_{0}(\boldsymbol{k})=\sum_{0}^{3}d_{i}(\boldsymbol{k})\Gamma_{i}
\end{eqnarray}
with $\Gamma_{i}=\sigma_{i}\otimes\tau_{1}$ for $i=1,2,3$,
and $\Gamma_{0}=\sigma_{0}\otimes\tau_{3}$. $\sigma_{i}$ and $\tau_{i}$
are Pauli matrices for the spin and orbital degrees of freedom,
respectively, $d_{0}=M_{0}+2B_{1}\left[1-\cos\left(k_{z}\right)\right]+2B_{2}\left[2-\cos\left(k_{x}\right)-\cos\left(k_{y}\right)\right]$, $d_{1}=A_{2}\sin\left(k_{x} \right)$, $d_{2}=A_{2}\sin\left(k_{y} \right)$, and $d_{3}=A_{1} \sin\left(k_{z} \right)$. 
The exchange field reads
\begin{eqnarray}
	H_{ex}=\begin{bmatrix}\vec{M_{A}}\cdot\vec{\sigma}\otimes\tau_{0} & 0\\
		0 & \vec{M_{B}}\cdot\vec{\sigma}\otimes\tau_{0}
	\end{bmatrix}.
\end{eqnarray}
For the antiferromagnetic MnBi$_2$Te$_4$, we introduce a sublayer index $i=A,B$ to describe the unit-cell doubling and characterize the magnetization in sublayer $i$ by $M_{i}=m(\cos\phi_{i}\,\sin\theta_{i},\sin\phi_{i}\sin\theta_{i},\cos\theta_{i})$ with angles $\phi_{i}$ and $\theta_{i}$. The A-type antiferromagnetic order is described by $(\theta_{A},\theta_{B})=(0,\pi)$ and $(\phi_{A},\phi_{B})=(0,0)$,
where
\begin{align}
	H_{ex}=\begin{bmatrix}m\,\sigma_{z}\otimes\tau_{0} & 0\\
		0 & -m\,\sigma_{z}\otimes\tau_{0}
	\end{bmatrix}.
\end{align}
The electric potential is 
$
V_E
= V(z) \, \sigma_0 \otimes \tau_0
$,
where $V(z)$ depicts the strength of the electric field. In the lattice model, the electric potential difference of neighboring layers is $V_z$. 

\begin{figure}[!tp]
	\centering
	\includegraphics[width=0.45\textwidth]{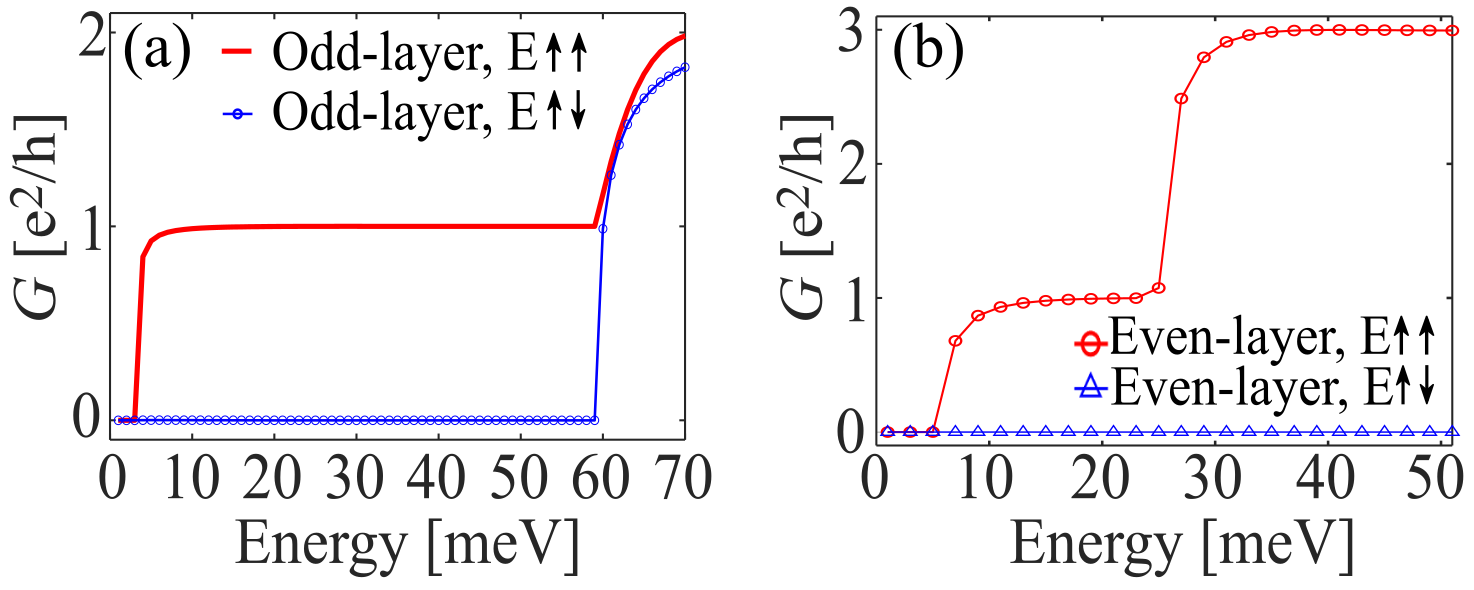}\\
	\caption{(a) Conductance $G$ as a function of Fermi energy for odd-layer thin films with $N_y$=20 and $N_z$=13. (b) Conductance $G$ for the mutimode case as a function of energy for even-layer thin films with $N_y$=50 and $N_z$=12. } \label{Fig:Intrindic_MTI_disorder}
\end{figure}

\section{More transport results} \label{sec:SM_transport}
\subsection{More results for intrinsic antiferromagnetic TIs}
Figure \ref{Fig:Intrindic_MTI_disorder}(a) shows that the magnetic topological transistor works well in the odd-layer thin film of antiferromagnetic TIs MnBi$_2$Te$_4$.
Figure \ref{Fig:Intrindic_MTI_disorder}(b) shows that when the size of $N_y$ increases, compared with Fig.~\ref{Fig:AFM_Normal_conductance}(b), more modes get involved and the transistor still works well. 

\subsection{Scattering matrix theory}
\begin{figure}[!tp]
	\centering
	\includegraphics[width=0.45\textwidth]{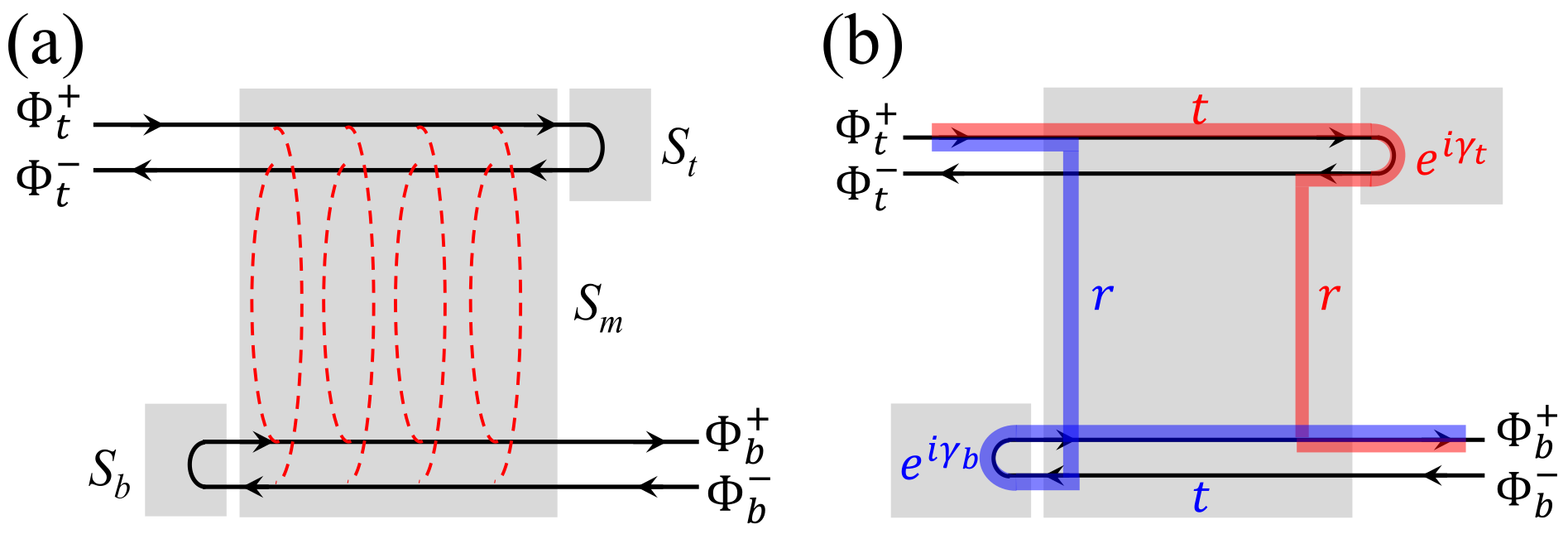}\\
	\caption{(a) Schematic of the scattering process for the ``off" state of the magnetic topological transistor with $\Phi_t^{\pm}$ and $\Phi_b^{\pm}$ the top and bottom right-moving and left-moving modes, respectively. $S_m$ and $S_{t/b}$ are the scattering matrices. (b)Two scattering paths for the transmission $T(E)$. Quantum interference occurs between two paths (red and blue thick lines).} \label{Fig:Scattering}
\end{figure}

By combining the scattering matrices $S_t$, $S_m$ and $S_b$ illustrated in Fig.~\ref{Fig:Scattering} (a), we obtain the scattering matrix connecting the states $\Phi_t^+,\,\Phi_b^-,\,\,\Phi_b^+,\,\Phi_t^-$. We solve the scattering problem using the one-dimensional effective Hamiltonian. As shown in Fig.~\ref{Fig:Scattering}(a), the right-moving mode in the top layer is reflected at the right edge, while the left-moving mode is reflected at the left edge. These scattering processes are described by $S_{t}$ and $S_{b}$. Namely, $S_{t}$ and $S_{b}$ model the barriers in the middle region.

The scattering matrix $S_m$ is obtained by solving two copies of the scattering problem: one is between $\Phi_t^+,\,\Phi_b^-$ and the other is between $\Phi_b^+,\,\Phi_t^-$, each with the Hamiltonian $vk_x\sigma_z + m \sigma_x$. Consequently, $S_m$ is
	\begin{equation}
		\left[\begin{array}{c}
			\Phi_t^+ \\
			\Phi_b^- \\
			\Phi_t^- \\
			\Phi_b^+
		\end{array}\right]_\text{out}
		=\left[\begin{array}{cccc}
			t & -r & 0 & 0 \\
			r & t & 0 & 0 \\
			0 & 0 & t & -r\\
			0 & 0 & r & t
		\end{array}\right]
		\left[\begin{array}{c}
			\Phi_t^+ \\
			\Phi_b^- \\
			\Phi_t^- \\
			\Phi_b^+ 
		\end{array}\right]_\text{in},
	\end{equation}
where $r$ and $t$ are the scattering amplitudes.
The scattering matrix $S_t$ connects only the scattering states $\Phi_t^\pm$ with $S_{t} = e^{i\gamma_{t}}$. Similarly, the scattering states on the bottom layer $\Phi_b^\pm$ are connected by $S_{b} = e^{i\gamma_{b}}$.
	
Combining the partial scattering matrices $S_t$, $S_m$, and $S_b$, the full scattering matrix $S$ becomes
	\begin{equation}
		\left[\begin{array}{c}
			\Phi_t^- \\
			\Phi_b^+
		\end{array}\right]_\text{out}
		=\left[\begin{array}{cccc}
			e^{i\gamma_t}t^2 - e^{i\gamma_b}r^2 && -(e^{i\gamma_t}+e^{i\gamma_b})rt \\
			(e^{i\gamma_t}+e^{i\gamma_b})rt && e^{i\gamma_b}t^2 - e^{i\gamma_t}r^2\\
		\end{array}\right]
		\left[\begin{array}{c}
			\Phi_t^+ \\
			\Phi_b^-
		\end{array}\right]_\text{in}. \nonumber
	\end{equation}

The transmission probability $T(E)$ for the incoming state $\Phi_t^+$ on the top layer to be transmitted to the outgoing mode $\Phi_b^+$ on the bottom layer reads
	\begin{eqnarray} \label{Eq:T(E)}
		T(E) 
		&=& |(e^{i \gamma_t} + e^{i \gamma_b} )rt|^2 
		= 4|rt|^2\cos^2[\Delta\gamma/2]
	\end{eqnarray}
with $\Delta\gamma = \gamma_t - \gamma_b$. Therefore, $T(E)$ can be understood in terms of quantum interference illustrated in Fig.~\ref{Fig:Scattering}(b).
 
We solve the scattering problem to obtain the phase difference $\Delta\gamma=\gamma_t-\gamma_b$. The reflection phase at the top layer $\gamma_t$ is obtained by solving $\chi_{t,I} + r_t \chi_{t,R} = \chi_{t,E}$. In particular, the reflection phase is
\begin{equation}
		\gamma_t = \pi + \arg\{\chi_{t,R}^\dagger\bar{\chi}_{t,E}\bar{\chi}_{t,E}^\dagger\chi_{t,I}\},
\end{equation}where the spin state $\bar{\chi}$ refers to the oppositely polarized state of $\chi$.
The second term can be interpreted as a geometric phase calculated by integrating the Berry connection along the geodesic connecting three spin states $\chi_{t,I}$, $\bar{\chi}_{t,E}$, and $\chi_{t,R}$ on the Bloch sphere. Similarly, for the bottom layer, we solve the scattering problem $\chi_{b,I} + r_b \chi_{b,R} = \chi_{b,E}$. The reflection phase $\gamma_b$ is
	\begin{equation}
		\gamma_b = \pi + \arg\{\chi_{b,R}^\dagger\bar{\chi}_{b,E}\bar{\chi}_{b,E}^\dagger\chi_{b,I}\}.
	\end{equation}
In total,
	\begin{equation}
		\Delta\gamma = \arg\{\chi_{t,R}^\dagger\bar{\chi}_{t,E}\bar{\chi}_{t,E}^\dagger\chi_{t,I} \chi_{b,I}^\dagger\bar{\chi}_{b,E}\bar{\chi}_{b,E}^\dagger\chi_{b,R}\}.
	\end{equation}
Finally, we obtain the quantum phase 
	\begin{equation}
		\Delta\gamma=\left\{
		\begin{array}{cc}
			0 & \, \text{odd layers}, \\
			\pi & \, \text{even layers}.
		\end{array}\right.
	\end{equation}
Hence, the transmission probability is
	\begin{equation}
		T(E) =
		\left\{
		\begin{array}{cc}
			4|rt|^2 & \, \text{odd layers}, \\
			0 & \, \text{even layers}.
		\end{array}\right. \nonumber
	\end{equation}
The interference at odd and even layers occurs constructively and destructively, respectively. Hence, the transmission is finite for the odd layers. On the contrary, for the even layers, the destructive interference suppresses the transport.


\begin{thebibliography}{68}%
\makeatletter
\providecommand \@ifxundefined [1]{%
 \@ifx{#1\undefined}
}%
\providecommand \@ifnum [1]{%
 \ifnum #1\expandafter \@firstoftwo
 \else \expandafter \@secondoftwo
 \fi
}%
\providecommand \@ifx [1]{%
 \ifx #1\expandafter \@firstoftwo
 \else \expandafter \@secondoftwo
 \fi
}%
\providecommand \natexlab [1]{#1}%
\providecommand \enquote  [1]{``#1''}%
\providecommand \bibnamefont  [1]{#1}%
\providecommand \bibfnamefont [1]{#1}%
\providecommand \citenamefont [1]{#1}%
\providecommand \href@noop [0]{\@secondoftwo}%
\providecommand \href [0]{\begingroup \@sanitize@url \@href}%
\providecommand \@href[1]{\@@startlink{#1}\@@href}%
\providecommand \@@href[1]{\endgroup#1\@@endlink}%
\providecommand \@sanitize@url [0]{\catcode `\\12\catcode `\$12\catcode
  `\&12\catcode `\#12\catcode `\^12\catcode `\_12\catcode `\%12\relax}%
\providecommand \@@startlink[1]{}%
\providecommand \@@endlink[0]{}%
\providecommand \url  [0]{\begingroup\@sanitize@url \@url }%
\providecommand \@url [1]{\endgroup\@href {#1}{\urlprefix }}%
\providecommand \urlprefix  [0]{URL }%
\providecommand \Eprint [0]{\href }%
\providecommand \doibase [0]{http://dx.doi.org/}%
\providecommand \selectlanguage [0]{\@gobble}%
\providecommand \bibinfo  [0]{\@secondoftwo}%
\providecommand \bibfield  [0]{\@secondoftwo}%
\providecommand \translation [1]{[#1]}%
\providecommand \BibitemOpen [0]{}%
\providecommand \bibitemStop [0]{}%
\providecommand \bibitemNoStop [0]{.\EOS\space}%
\providecommand \EOS [0]{\spacefactor3000\relax}%
\providecommand \BibitemShut  [1]{\csname bibitem#1\endcsname}%
\let\auto@bib@innerbib\@empty
\bibitem [{\citenamefont {Moore}(2010)}]{Moore10NBirth}%
  \BibitemOpen
  \bibfield  {author} {\bibinfo {author} {\bibfnamefont {J.~E.}\ \bibnamefont
  {Moore}},\ }\bibfield  {title} {\enquote {\bibinfo {title} {The birth of
  topological insulators}}, }\href
  {https://www.nature.com/articles/nature08916} {\bibfield  {journal} {\bibinfo
   {journal} {Nature}\ }\textbf {\bibinfo {volume} {464}},\ \bibinfo {pages}
  {194} (\bibinfo {year} {2010})}\BibitemShut {NoStop}%
\bibitem [{\citenamefont {Hasan}\ and\ \citenamefont
  {Kane}(2010)}]{Hasan10RMPColloquium}%
  \BibitemOpen
  \bibfield  {author} {\bibinfo {author} {\bibfnamefont {M.~Z.}\ \bibnamefont
  {Hasan}}\ and\ \bibinfo {author} {\bibfnamefont {C.~L.}\ \bibnamefont
  {Kane}},\ }\bibfield  {title} {\enquote {\bibinfo {title} {Colloquium:
  {{Topological}} insulators}}, }\href
  {https://link.aps.org/doi/10.1103/RevModPhys.82.3045} {\bibfield  {journal}
  {\bibinfo  {journal} {Rev. Mod. Phys.}\ }\textbf {\bibinfo {volume} {82}},\
  \bibinfo {pages} {3045} (\bibinfo {year} {2010})}\BibitemShut {NoStop}%
\bibitem [{\citenamefont {Qi}\ and\ \citenamefont
  {Zhang}(2011)}]{Qi11RMPTopological}%
  \BibitemOpen
  \bibfield  {author} {\bibinfo {author} {\bibfnamefont {X.-L.}\ \bibnamefont
  {Qi}}\ and\ \bibinfo {author} {\bibfnamefont {S.-C.}\ \bibnamefont {Zhang}},\
  }\bibfield  {title} {\enquote {\bibinfo {title} {Topological insulators and
  superconductors}}, }\href
  {https://link.aps.org/doi/10.1103/RevModPhys.83.1057} {\bibfield  {journal}
  {\bibinfo  {journal} {Rev. Mod. Phys.}\ }\textbf {\bibinfo {volume} {83}},\
  \bibinfo {pages} {1057} (\bibinfo {year} {2011})}\BibitemShut {NoStop}%
\bibitem [{\citenamefont {Xue}(2011)}]{Xue11NNTopological}%
  \BibitemOpen
  \bibfield  {author} {\bibinfo {author} {\bibfnamefont {Q.-K.}\ \bibnamefont
  {Xue}},\ }\bibfield  {title} {\enquote {\bibinfo {title} {A topological twist
  for transistors}}, }\href {\doibase 10.1038/nnano.2011.47} {\bibfield
  {journal} {\bibinfo  {journal} {Nat. Nanotechnol.}\ }\textbf {\bibinfo
  {volume} {6}},\ \bibinfo {pages} {197} (\bibinfo {year} {2011})}\BibitemShut
  {NoStop}%
\bibitem [{\citenamefont {Xiu}\ \emph {et~al.}(2011)\citenamefont {Xiu},
  \citenamefont {He}, \citenamefont {Wang}, \citenamefont {Cheng},
  \citenamefont {Chang}, \citenamefont {Lang}, \citenamefont {Huang},
  \citenamefont {Kou}, \citenamefont {Zhou}, \citenamefont {Jiang},
  \citenamefont {Chen}, \citenamefont {Zou}, \citenamefont {Shailos},\ and\
  \citenamefont {Wang}}]{Xiu11NNManipulating}%
  \BibitemOpen
  \bibfield  {author} {\bibinfo {author} {\bibfnamefont {F.}~\bibnamefont
  {Xiu}},  \emph {et~al.},\ }\bibfield  {title} {\enquote {\bibinfo {title}
  {Manipulating surface states in topological insulator nanoribbons}}, }\href
  {\doibase 10.1038/nnano.2011.19} {\bibfield  {journal} {\bibinfo  {journal}
  {Nat. Nanotechnol.}\ }\textbf {\bibinfo {volume} {6}},\ \bibinfo {pages}
  {216} (\bibinfo {year} {2011})}\BibitemShut {NoStop}%
\bibitem [{\citenamefont {Wray}(2012)}]{Wray12NPTopological}%
  \BibitemOpen
  \bibfield  {author} {\bibinfo {author} {\bibfnamefont {L.~A.}\ \bibnamefont
  {Wray}},\ }\bibfield  {title} {\enquote {\bibinfo {title} {Topological
  transistor}}, }\href {\doibase 10.1038/nphys2410} {\bibfield  {journal}
  {\bibinfo  {journal} {Nat. Phys.}\ }\textbf {\bibinfo {volume} {8}},\
  \bibinfo {pages} {705} (\bibinfo {year} {2012})}\BibitemShut {NoStop}%
\bibitem [{\citenamefont {Checkelsky}\ \emph {et~al.}(2012)\citenamefont
  {Checkelsky}, \citenamefont {Ye}, \citenamefont {Onose}, \citenamefont
  {Iwasa},\ and\ \citenamefont {Tokura}}]{Checkelsky12NPDiracfermionmediated}%
  \BibitemOpen
  \bibfield  {author} {\bibinfo {author} {\bibfnamefont {J.~G.}\ \bibnamefont
  {Checkelsky}}, \bibinfo {author} {\bibfnamefont {J.}~\bibnamefont {Ye}},
  \bibinfo {author} {\bibfnamefont {Y.}~\bibnamefont {Onose}}, \bibinfo
  {author} {\bibfnamefont {Y.}~\bibnamefont {Iwasa}}, \ and\ \bibinfo {author}
  {\bibfnamefont {Y.}~\bibnamefont {Tokura}},\ }\bibfield  {title} {\enquote
  {\bibinfo {title} {Dirac-fermion-mediated ferromagnetism in a topological
  insulator}}, }\href {\doibase 10.1038/nphys2388} {\bibfield  {journal}
  {\bibinfo  {journal} {Nat. Phys.}\ }\textbf {\bibinfo {volume} {8}},\
  \bibinfo {pages} {729} (\bibinfo {year} {2012})}\BibitemShut {NoStop}%
\bibitem [{\citenamefont {Yu}\ \emph {et~al.}(2010)\citenamefont {Yu},
  \citenamefont {Zhang}, \citenamefont {Zhang}, \citenamefont {Zhang},
  \citenamefont {Dai},\ and\ \citenamefont {Fang}}]{Yu10SQuantized}%
  \BibitemOpen
  \bibfield  {author} {\bibinfo {author} {\bibfnamefont {R.}~\bibnamefont
  {Yu}}, \bibinfo {author} {\bibfnamefont {W.}~\bibnamefont {Zhang}}, \bibinfo
  {author} {\bibfnamefont {H.-J.}\ \bibnamefont {Zhang}}, \bibinfo {author}
  {\bibfnamefont {S.-C.}\ \bibnamefont {Zhang}}, \bibinfo {author}
  {\bibfnamefont {X.}~\bibnamefont {Dai}}, \ and\ \bibinfo {author}
  {\bibfnamefont {Z.}~\bibnamefont {Fang}},\ }\bibfield  {title} {\enquote
  {\bibinfo {title} {Quantized {{Anomalous Hall Effect}} in {{Magnetic
  Topological Insulators}}}}, }\href
  {https://science.sciencemag.org/content/329/5987/61} {\bibfield  {journal}
  {\bibinfo  {journal} {Science}\ }\textbf {\bibinfo {volume} {329}},\ \bibinfo
  {pages} {61} (\bibinfo {year} {2010})}\BibitemShut {NoStop}%
\bibitem [{\citenamefont {Michetti}\ and\ \citenamefont
  {Trauzettel}(2013)}]{Michetti13APLDevices}%
  \BibitemOpen
  \bibfield  {author} {\bibinfo {author} {\bibfnamefont {P.}~\bibnamefont
  {Michetti}}\ and\ \bibinfo {author} {\bibfnamefont {B.}~\bibnamefont
  {Trauzettel}},\ }\bibfield  {title} {\enquote {\bibinfo {title} {Devices with
  electrically tunable topological insulating phases}}, }\href {\doibase
  10.1063/1.4792275} {\bibfield  {journal} {\bibinfo  {journal} {Appl. Phys.
  Lett.}\ }\textbf {\bibinfo {volume} {102}},\ \bibinfo {pages} {063503}
  (\bibinfo {year} {2013})}\BibitemShut {NoStop}%
\bibitem [{\citenamefont {Qian}\ \emph {et~al.}(2014)\citenamefont {Qian},
  \citenamefont {Liu}, \citenamefont {Fu},\ and\ \citenamefont
  {Li}}]{Qian14SQuantum}%
  \BibitemOpen
  \bibfield  {author} {\bibinfo {author} {\bibfnamefont {X.}~\bibnamefont
  {Qian}}, \bibinfo {author} {\bibfnamefont {J.}~\bibnamefont {Liu}}, \bibinfo
  {author} {\bibfnamefont {L.}~\bibnamefont {Fu}}, \ and\ \bibinfo {author}
  {\bibfnamefont {J.}~\bibnamefont {Li}},\ }\bibfield  {title} {\enquote
  {\bibinfo {title} {Quantum spin {{Hall}} effect in two-dimensional transition
  metal dichalcogenides}}, }\href {\doibase 10.1126/science.1256815} {\bibfield
   {journal} {\bibinfo  {journal} {Science}\ }\textbf {\bibinfo {volume}
  {346}},\ \bibinfo {pages} {1344} (\bibinfo {year} {2014})}\BibitemShut
  {NoStop}%
\bibitem [{\citenamefont {Liu}\ \emph {et~al.}(2014)\citenamefont {Liu},
  \citenamefont {Hsieh}, \citenamefont {Wei}, \citenamefont {Duan},
  \citenamefont {Moodera},\ and\ \citenamefont {Fu}}]{Liu14NMSpinfiltered}%
  \BibitemOpen
  \bibfield  {author} {\bibinfo {author} {\bibfnamefont {J.}~\bibnamefont
  {Liu}}, \bibinfo {author} {\bibfnamefont {T.~H.}\ \bibnamefont {Hsieh}},
  \bibinfo {author} {\bibfnamefont {P.}~\bibnamefont {Wei}}, \bibinfo {author}
  {\bibfnamefont {W.}~\bibnamefont {Duan}}, \bibinfo {author} {\bibfnamefont
  {J.}~\bibnamefont {Moodera}}, \ and\ \bibinfo {author} {\bibfnamefont
  {L.}~\bibnamefont {Fu}},\ }\bibfield  {title} {\enquote {\bibinfo {title}
  {Spin-filtered edge states with an electrically tunable gap in a
  two-dimensional topological crystalline insulator}}, }\href {\doibase
  10.1038/nmat3828} {\bibfield  {journal} {\bibinfo  {journal} {Nat. Mater.}\
  }\textbf {\bibinfo {volume} {13}},\ \bibinfo {pages} {178} (\bibinfo {year}
  {2014})}\BibitemShut {NoStop}%
\bibitem [{\citenamefont {Wang}\ \emph {et~al.}(2015)\citenamefont {Wang},
  \citenamefont {Lian},\ and\ \citenamefont {Zhang}}]{Wang15PRLElectrically}%
  \BibitemOpen
  \bibfield  {author} {\bibinfo {author} {\bibfnamefont {J.}~\bibnamefont
  {Wang}}, \bibinfo {author} {\bibfnamefont {B.}~\bibnamefont {Lian}}, \ and\
  \bibinfo {author} {\bibfnamefont {S.-C.}\ \bibnamefont {Zhang}},\ }\bibfield
  {title} {\enquote {\bibinfo {title} {Electrically {{Tunable Magnetism}} in
  {{Magnetic Topological Insulators}}}}, }\href {\doibase
  10.1103/PhysRevLett.115.036805} {\bibfield  {journal} {\bibinfo  {journal}
  {Phys. Rev. Lett.}\ }\textbf {\bibinfo {volume} {115}},\ \bibinfo {pages}
  {036805} (\bibinfo {year} {2015})}\BibitemShut {NoStop}%
\bibitem [{\citenamefont {Liu}\ \emph {et~al.}(2015)\citenamefont {Liu},
  \citenamefont {Zhang}, \citenamefont {Abdalla}, \citenamefont {Fazzio},\ and\
  \citenamefont {Zunger}}]{Liu15NLSwitching}%
  \BibitemOpen
  \bibfield  {author} {\bibinfo {author} {\bibfnamefont {Q.}~\bibnamefont
  {Liu}}, \bibinfo {author} {\bibfnamefont {X.}~\bibnamefont {Zhang}}, \bibinfo
  {author} {\bibfnamefont {L.~B.}\ \bibnamefont {Abdalla}}, \bibinfo {author}
  {\bibfnamefont {A.}~\bibnamefont {Fazzio}}, \ and\ \bibinfo {author}
  {\bibfnamefont {A.}~\bibnamefont {Zunger}},\ }\bibfield  {title} {\enquote
  {\bibinfo {title} {Switching a {{Normal Insulator}} into a {{Topological
  Insulator}} via {{Electric Field}} with {{Application}} to {{Phosphorene}}}},
  }\href {\doibase 10.1021/nl5043769} {\bibfield  {journal} {\bibinfo
  {journal} {Nano Lett.}\ }\textbf {\bibinfo {volume} {15}},\ \bibinfo {pages}
  {1222} (\bibinfo {year} {2015})}\BibitemShut {NoStop}%
\bibitem [{\citenamefont {Collins}\ \emph {et~al.}(2018)\citenamefont
  {Collins}, \citenamefont {Tadich}, \citenamefont {Wu}, \citenamefont {Gomes},
  \citenamefont {Rodrigues}, \citenamefont {Liu}, \citenamefont {Hellerstedt},
  \citenamefont {Ryu}, \citenamefont {Tang}, \citenamefont {Mo}, \citenamefont
  {Adam}, \citenamefont {Yang}, \citenamefont {Fuhrer},\ and\ \citenamefont
  {Edmonds}}]{Collins18NElectricfieldtuned}%
  \BibitemOpen
  \bibfield  {author} {\bibinfo {author} {\bibfnamefont {J.~L.}\ \bibnamefont
  {Collins}},  \emph {et~al.},\ }\bibfield  {title} {\enquote {\bibinfo {title}
  {Electric-field-tuned topological phase transition in ultrathin
  {{Na}}$_3${{Bi}}}}, }\href {\doibase 10.1038/s41586-018-0788-5} {\bibfield
  {journal} {\bibinfo  {journal} {Nature}\ }\textbf {\bibinfo {volume} {564}},\
  \bibinfo {pages} {390} (\bibinfo {year} {2018})}\BibitemShut {NoStop}%
\bibitem [{\citenamefont {Zhang}\ \emph {et~al.}(2019)\citenamefont {Zhang},
  \citenamefont {Shi}, \citenamefont {Zhu}, \citenamefont {Xing}, \citenamefont
  {Zhang},\ and\ \citenamefont {Wang}}]{Zhang19PRLTopological}%
  \BibitemOpen
  \bibfield  {author} {\bibinfo {author} {\bibfnamefont {D.}~\bibnamefont
  {Zhang}}, \bibinfo {author} {\bibfnamefont {M.}~\bibnamefont {Shi}}, \bibinfo
  {author} {\bibfnamefont {T.}~\bibnamefont {Zhu}}, \bibinfo {author}
  {\bibfnamefont {D.}~\bibnamefont {Xing}}, \bibinfo {author} {\bibfnamefont
  {H.}~\bibnamefont {Zhang}}, \ and\ \bibinfo {author} {\bibfnamefont
  {J.}~\bibnamefont {Wang}},\ }\bibfield  {title} {\enquote {\bibinfo {title}
  {Topological {{Axion States}} in the {{Magnetic Insulator MnBi$_2$Te$_4$}}
  with the {{Quantized Magnetoelectric Effect}}}}, }\href
  {https://link.aps.org/doi/10.1103/PhysRevLett.122.206401} {\bibfield
  {journal} {\bibinfo  {journal} {Phys. Rev. Lett.}\ }\textbf {\bibinfo
  {volume} {122}},\ \bibinfo {pages} {206401} (\bibinfo {year}
  {2019})}\BibitemShut {NoStop}%
\bibitem [{\citenamefont {Otrokov}\ \emph
  {et~al.}(2019{\natexlab{a}})\citenamefont {Otrokov}, \citenamefont
  {Klimovskikh}, \citenamefont {Bentmann}, \citenamefont {Estyunin},
  \citenamefont {Zeugner}, \citenamefont {Aliev}, \citenamefont {Ga{\ss}},
  \citenamefont {Wolter}, \citenamefont {Koroleva}, \citenamefont {Shikin},
  \citenamefont {{Blanco-Rey}}, \citenamefont {Hoffmann}, \citenamefont
  {Rusinov}, \citenamefont {Vyazovskaya}, \citenamefont {Eremeev},
  \citenamefont {Koroteev}, \citenamefont {Kuznetsov}, \citenamefont {Freyse},
  \citenamefont {{S{\'a}nchez-Barriga}}, \citenamefont {Amiraslanov},
  \citenamefont {Babanly}, \citenamefont {Mamedov}, \citenamefont {Abdullayev},
  \citenamefont {Zverev}, \citenamefont {Alfonsov}, \citenamefont {Kataev},
  \citenamefont {B{\"u}chner}, \citenamefont {Schwier}, \citenamefont {Kumar},
  \citenamefont {Kimura}, \citenamefont {Petaccia}, \citenamefont {Di~Santo},
  \citenamefont {Vidal}, \citenamefont {Schatz}, \citenamefont {Ki{\ss}ner},
  \citenamefont {{\"U}nzelmann}, \citenamefont {Min}, \citenamefont {Moser},
  \citenamefont {Peixoto}, \citenamefont {Reinert}, \citenamefont {Ernst},
  \citenamefont {Echenique}, \citenamefont {Isaeva},\ and\ \citenamefont
  {Chulkov}}]{Otrokov19NPrediction}%
  \BibitemOpen
  \bibfield  {author} {\bibinfo {author} {\bibfnamefont {M.~M.}\ \bibnamefont
  {Otrokov}},  \emph {et~al.},\ }\bibfield  {title} {\enquote {\bibinfo {title}
  {Prediction and observation of an antiferromagnetic topological insulator}},
  }\href {\doibase 10.1038/s41586-019-1840-9} {\bibfield  {journal} {\bibinfo
  {journal} {Nature}\ }\textbf {\bibinfo {volume} {576}},\ \bibinfo {pages}
  {416} (\bibinfo {year} {2019}{\natexlab{a}})}\BibitemShut {NoStop}%
\bibitem [{\citenamefont {Rienks}\ \emph {et~al.}(2019)\citenamefont {Rienks},
  \citenamefont {Wimmer}, \citenamefont {{S{\'a}nchez-Barriga}}, \citenamefont
  {Caha}, \citenamefont {Mandal}, \citenamefont {R{\r{u}}{\v z}i{\v c}ka},
  \citenamefont {Ney}, \citenamefont {Steiner}, \citenamefont {Volobuev},
  \citenamefont {Groiss}, \citenamefont {Albu}, \citenamefont {Kothleitner},
  \citenamefont {Michali{\v c}ka}, \citenamefont {Khan}, \citenamefont
  {Min{\'a}r}, \citenamefont {Ebert}, \citenamefont {Bauer}, \citenamefont
  {Freyse}, \citenamefont {Varykhalov}, \citenamefont {Rader},\ and\
  \citenamefont {Springholz}}]{Rienks19NLarge}%
  \BibitemOpen
  \bibfield  {author} {\bibinfo {author} {\bibfnamefont {E.~D.~L.}\
  \bibnamefont {Rienks}},  \emph {et~al.},\ }\bibfield  {title} {\enquote
  {\bibinfo {title} {Large magnetic gap at the {{Dirac}} point in
  {{Bi}}$_2${{Te}}$_3$/{{MnBi}}$_2${{Te}}$_4$ heterostructures}}, }\href
  {https://www.nature.com/articles/s41586-019-1826-7} {\bibfield  {journal}
  {\bibinfo  {journal} {Nature}\ }\textbf {\bibinfo {volume} {576}},\ \bibinfo
  {pages} {423} (\bibinfo {year} {2019})}\BibitemShut {NoStop}%
\bibitem [{\citenamefont {Gong}\ \emph {et~al.}(2019)\citenamefont {Gong},
  \citenamefont {Guo}, \citenamefont {Li}, \citenamefont {Zhu}, \citenamefont
  {Liao}, \citenamefont {Liu}, \citenamefont {Zhang}, \citenamefont {Gu},
  \citenamefont {Tang}, \citenamefont {Feng}, \citenamefont {Zhang},
  \citenamefont {Li}, \citenamefont {Song}, \citenamefont {Wang}, \citenamefont
  {Yu}, \citenamefont {Chen}, \citenamefont {Wang}, \citenamefont {Yao},
  \citenamefont {Duan}, \citenamefont {Xu}, \citenamefont {Zhang},
  \citenamefont {Ma}, \citenamefont {Xue},\ and\ \citenamefont
  {He}}]{Gong19CPLExperimental}%
  \BibitemOpen
  \bibfield  {author} {\bibinfo {author} {\bibfnamefont {Y.}~\bibnamefont
  {Gong}},  \emph {et~al.},\ }\bibfield  {title} {\enquote {\bibinfo {title}
  {Experimental {{Realization}} of an {{Intrinsic Magnetic Topological
  Insulator}}}}, }\href {https://doi.org/10.1088/0256-307x/36/7/076801}
  {\bibfield  {journal} {\bibinfo  {journal} {Chinese Phys. Lett.}\ }\textbf
  {\bibinfo {volume} {36}},\ \bibinfo {pages} {076801} (\bibinfo {year}
  {2019})}\BibitemShut {NoStop}%
\bibitem [{\citenamefont {Li}\ \emph {et~al.}(2019{\natexlab{a}})\citenamefont
  {Li}, \citenamefont {Li}, \citenamefont {Du}, \citenamefont {Wang},
  \citenamefont {Gu}, \citenamefont {Zhang}, \citenamefont {He}, \citenamefont
  {Duan},\ and\ \citenamefont {Xu}}]{Li19SAIntrinsic}%
  \BibitemOpen
  \bibfield  {author} {\bibinfo {author} {\bibfnamefont {J.}~\bibnamefont
  {Li}}, \bibinfo {author} {\bibfnamefont {Y.}~\bibnamefont {Li}}, \bibinfo
  {author} {\bibfnamefont {S.}~\bibnamefont {Du}}, \bibinfo {author}
  {\bibfnamefont {Z.}~\bibnamefont {Wang}}, \bibinfo {author} {\bibfnamefont
  {B.-L.}\ \bibnamefont {Gu}}, \bibinfo {author} {\bibfnamefont {S.-C.}\
  \bibnamefont {Zhang}}, \bibinfo {author} {\bibfnamefont {K.}~\bibnamefont
  {He}}, \bibinfo {author} {\bibfnamefont {W.}~\bibnamefont {Duan}}, \ and\
  \bibinfo {author} {\bibfnamefont {Y.}~\bibnamefont {Xu}},\ }\bibfield
  {title} {\enquote {\bibinfo {title} {Intrinsic magnetic topological
  insulators in van der {{Waals}} layered {{MnBi}}$_2${{Te}}$_4$-family
  materials}}, }\href {https://advances.sciencemag.org/content/5/6/eaaw5685}
  {\bibfield  {journal} {\bibinfo  {journal} {Sci. Adv.}\ }\textbf {\bibinfo
  {volume} {5}},\ \bibinfo {pages} {eaaw5685} (\bibinfo {year}
  {2019}{\natexlab{a}})}\BibitemShut {NoStop}%
\bibitem [{\citenamefont {Otrokov}\ \emph
  {et~al.}(2019{\natexlab{b}})\citenamefont {Otrokov}, \citenamefont {Rusinov},
  \citenamefont {{Blanco-Rey}}, \citenamefont {Hoffmann}, \citenamefont
  {Vyazovskaya}, \citenamefont {Eremeev}, \citenamefont {Ernst}, \citenamefont
  {Echenique}, \citenamefont {Arnau},\ and\ \citenamefont
  {Chulkov}}]{Otrokov19PRLUnique}%
  \BibitemOpen
  \bibfield  {author} {\bibinfo {author} {\bibfnamefont {M.~M.}\ \bibnamefont
  {Otrokov}}, \bibinfo {author} {\bibfnamefont {I.~P.}\ \bibnamefont
  {Rusinov}}, \bibinfo {author} {\bibfnamefont {M.}~\bibnamefont
  {{Blanco-Rey}}}, \bibinfo {author} {\bibfnamefont {M.}~\bibnamefont
  {Hoffmann}}, \bibinfo {author} {\bibfnamefont {A.~Y.}\ \bibnamefont
  {Vyazovskaya}}, \bibinfo {author} {\bibfnamefont {S.~V.}\ \bibnamefont
  {Eremeev}}, \bibinfo {author} {\bibfnamefont {A.}~\bibnamefont {Ernst}},
  \bibinfo {author} {\bibfnamefont {P.~M.}\ \bibnamefont {Echenique}}, \bibinfo
  {author} {\bibfnamefont {A.}~\bibnamefont {Arnau}}, \ and\ \bibinfo {author}
  {\bibfnamefont {E.~V.}\ \bibnamefont {Chulkov}},\ }\bibfield  {title}
  {\enquote {\bibinfo {title} {Unique {{Thickness-Dependent Properties}} of the
  van der {{Waals Interlayer Antiferromagnet MnBi}}$_2${{Te}}$_4$ {{Films}}}},
  }\href {https://link.aps.org/doi/10.1103/PhysRevLett.122.107202} {\bibfield
  {journal} {\bibinfo  {journal} {Phys. Rev. Lett.}\ }\textbf {\bibinfo
  {volume} {122}},\ \bibinfo {pages} {107202} (\bibinfo {year}
  {2019}{\natexlab{b}})}\BibitemShut {NoStop}%
\bibitem [{\citenamefont {Sun}\ \emph {et~al.}(2019)\citenamefont {Sun},
  \citenamefont {Xia}, \citenamefont {Chen}, \citenamefont {Zhang},
  \citenamefont {Liu}, \citenamefont {Yao}, \citenamefont {Tang}, \citenamefont
  {Zhao}, \citenamefont {Xu},\ and\ \citenamefont {Liu}}]{Sun19PRLRational}%
  \BibitemOpen
  \bibfield  {author} {\bibinfo {author} {\bibfnamefont {H.}~\bibnamefont
  {Sun}}, \bibinfo {author} {\bibfnamefont {B.}~\bibnamefont {Xia}}, \bibinfo
  {author} {\bibfnamefont {Z.}~\bibnamefont {Chen}}, \bibinfo {author}
  {\bibfnamefont {Y.}~\bibnamefont {Zhang}}, \bibinfo {author} {\bibfnamefont
  {P.}~\bibnamefont {Liu}}, \bibinfo {author} {\bibfnamefont {Q.}~\bibnamefont
  {Yao}}, \bibinfo {author} {\bibfnamefont {H.}~\bibnamefont {Tang}}, \bibinfo
  {author} {\bibfnamefont {Y.}~\bibnamefont {Zhao}}, \bibinfo {author}
  {\bibfnamefont {H.}~\bibnamefont {Xu}}, \ and\ \bibinfo {author}
  {\bibfnamefont {Q.}~\bibnamefont {Liu}},\ }\bibfield  {title} {\enquote
  {\bibinfo {title} {Rational {{Design Principles}} of the {{Quantum Anomalous
  Hall Effect}} in {{Superlatticelike Magnetic Topological Insulators}}}},
  }\href {\doibase 10.1103/PhysRevLett.123.096401} {\bibfield  {journal}
  {\bibinfo  {journal} {Phys. Rev. Lett.}\ }\textbf {\bibinfo {volume} {123}},\
  \bibinfo {pages} {096401} (\bibinfo {year} {2019})}\BibitemShut {NoStop}%
\bibitem [{\citenamefont {Lei}\ \emph {et~al.}(2020)\citenamefont {Lei},
  \citenamefont {Chen},\ and\ \citenamefont {MacDonald}}]{Lei20PNASMagnetized}%
  \BibitemOpen
  \bibfield  {author} {\bibinfo {author} {\bibfnamefont {C.}~\bibnamefont
  {Lei}}, \bibinfo {author} {\bibfnamefont {S.}~\bibnamefont {Chen}}, \ and\
  \bibinfo {author} {\bibfnamefont {A.~H.}\ \bibnamefont {MacDonald}},\
  }\bibfield  {title} {\enquote {\bibinfo {title} {Magnetized topological
  insulator multilayers}}, }\href {\doibase 10.1073/pnas.2014004117} {\bibfield
   {journal} {\bibinfo  {journal} {Proceedings of the National Academy of
  Sciences}\ }\textbf {\bibinfo {volume} {117}},\ \bibinfo {pages} {27224}
  (\bibinfo {year} {2020})}\BibitemShut {NoStop}%
\bibitem [{\citenamefont {Wang}\ \emph {et~al.}(2020)\citenamefont {Wang},
  \citenamefont {Wang}, \citenamefont {Yang}, \citenamefont {Shi},
  \citenamefont {Ruan}, \citenamefont {Xing}, \citenamefont {Wang},\ and\
  \citenamefont {Zhang}}]{Wang20PRBDynamical}%
  \BibitemOpen
  \bibfield  {author} {\bibinfo {author} {\bibfnamefont {H.}~\bibnamefont
  {Wang}}, \bibinfo {author} {\bibfnamefont {D.}~\bibnamefont {Wang}}, \bibinfo
  {author} {\bibfnamefont {Z.}~\bibnamefont {Yang}}, \bibinfo {author}
  {\bibfnamefont {M.}~\bibnamefont {Shi}}, \bibinfo {author} {\bibfnamefont
  {J.}~\bibnamefont {Ruan}}, \bibinfo {author} {\bibfnamefont {D.}~\bibnamefont
  {Xing}}, \bibinfo {author} {\bibfnamefont {J.}~\bibnamefont {Wang}}, \ and\
  \bibinfo {author} {\bibfnamefont {H.}~\bibnamefont {Zhang}},\ }\bibfield
  {title} {\enquote {\bibinfo {title} {Dynamical axion state with hidden
  pseudospin {{Chern}} numbers in {{MnBi}}$_2${{Te}}$_4$-based
  heterostructures}}, }\href {\doibase 10.1103/PhysRevB.101.081109} {\bibfield
  {journal} {\bibinfo  {journal} {Phys. Rev. B}\ }\textbf {\bibinfo {volume}
  {101}},\ \bibinfo {pages} {081109} (\bibinfo {year} {2020})}\BibitemShut
  {NoStop}%
\bibitem [{\citenamefont {Zhang}\ \emph {et~al.}(2020)\citenamefont {Zhang},
  \citenamefont {Wu},\ and\ \citenamefont {Das~Sarma}}]{Zhang20PRLMobius}%
  \BibitemOpen
  \bibfield  {author} {\bibinfo {author} {\bibfnamefont {R.-X.}\ \bibnamefont
  {Zhang}}, \bibinfo {author} {\bibfnamefont {F.}~\bibnamefont {Wu}}, \ and\
  \bibinfo {author} {\bibfnamefont {S.}~\bibnamefont {Das~Sarma}},\ }\bibfield
  {title} {\enquote {\bibinfo {title} {M\"obius {{Insulator}} and
  {{Higher-Order Topology}} in {{MnBi}}$_{2n}${{Te}}$_{3n+1}$}}, }\href
  {\doibase 10.1103/PhysRevLett.124.136407} {\bibfield  {journal} {\bibinfo
  {journal} {Phys. Rev. Lett.}\ }\textbf {\bibinfo {volume} {124}},\ \bibinfo
  {pages} {136407} (\bibinfo {year} {2020})}\BibitemShut {NoStop}%
\bibitem [{\citenamefont {Lian}\ \emph {et~al.}(2020)\citenamefont {Lian},
  \citenamefont {Liu}, \citenamefont {Zhang},\ and\ \citenamefont
  {Wang}}]{Lian20PRLFlat}%
  \BibitemOpen
  \bibfield  {author} {\bibinfo {author} {\bibfnamefont {B.}~\bibnamefont
  {Lian}}, \bibinfo {author} {\bibfnamefont {Z.}~\bibnamefont {Liu}}, \bibinfo
  {author} {\bibfnamefont {Y.}~\bibnamefont {Zhang}}, \ and\ \bibinfo {author}
  {\bibfnamefont {J.}~\bibnamefont {Wang}},\ }\bibfield  {title} {\enquote
  {\bibinfo {title} {Flat {{Chern Band}} from {{Twisted Bilayer
  MnBi}}$_2${{Te}}$_4$}}, }\href {\doibase 10.1103/PhysRevLett.124.126402}
  {\bibfield  {journal} {\bibinfo  {journal} {Phys. Rev. Lett.}\ }\textbf
  {\bibinfo {volume} {124}},\ \bibinfo {pages} {126402} (\bibinfo {year}
  {2020})}\BibitemShut {NoStop}%
\bibitem [{\citenamefont {Lei}\ and\ \citenamefont
  {MacDonald}(2021)}]{Lei21PRMGatetunable}%
  \BibitemOpen
  \bibfield  {author} {\bibinfo {author} {\bibfnamefont {C.}~\bibnamefont
  {Lei}}\ and\ \bibinfo {author} {\bibfnamefont {A.~H.}\ \bibnamefont
  {MacDonald}},\ }\bibfield  {title} {\enquote {\bibinfo {title} {Gate-tunable
  quantum anomalous {{Hall}} effects in {{MnBi}}$_2${{Te}}$_4$ thin films}},
  }\href {\doibase 10.1103/PhysRevMaterials.5.L051201} {\bibfield  {journal}
  {\bibinfo  {journal} {Phys. Rev. Materials}\ }\textbf {\bibinfo {volume}
  {5}},\ \bibinfo {pages} {L051201} (\bibinfo {year} {2021})}\BibitemShut
  {NoStop}%
\bibitem [{\citenamefont {Wei}\ \emph {et~al.}(2021)\citenamefont {Wei},
  \citenamefont {Zhu}, \citenamefont {Song},\ and\ \citenamefont
  {Chang}}]{Wei21PRBRenormalization}%
  \BibitemOpen
  \bibfield  {author} {\bibinfo {author} {\bibfnamefont {B.}~\bibnamefont
  {Wei}}, \bibinfo {author} {\bibfnamefont {J.-J.}\ \bibnamefont {Zhu}},
  \bibinfo {author} {\bibfnamefont {Y.}~\bibnamefont {Song}}, \ and\ \bibinfo
  {author} {\bibfnamefont {K.}~\bibnamefont {Chang}},\ }\bibfield  {title}
  {\enquote {\bibinfo {title} {Renormalization of gapped magnon excitation in
  monolayer {{MnBi}}$_2${{Te}}$_4$ by magnon-magnon interaction}}, }\href
  {\doibase 10.1103/PhysRevB.104.174436} {\bibfield  {journal} {\bibinfo
  {journal} {Phys. Rev. B}\ }\textbf {\bibinfo {volume} {104}},\ \bibinfo
  {pages} {174436} (\bibinfo {year} {2021})}\BibitemShut {NoStop}%
\bibitem [{\citenamefont {Varnava}\ \emph {et~al.}(2021)\citenamefont
  {Varnava}, \citenamefont {Wilson}, \citenamefont {Pixley},\ and\
  \citenamefont {Vanderbilt}}]{Varnava21NCControllable}%
  \BibitemOpen
  \bibfield  {author} {\bibinfo {author} {\bibfnamefont {N.}~\bibnamefont
  {Varnava}}, \bibinfo {author} {\bibfnamefont {J.~H.}\ \bibnamefont {Wilson}},
  \bibinfo {author} {\bibfnamefont {J.~H.}\ \bibnamefont {Pixley}}, \ and\
  \bibinfo {author} {\bibfnamefont {D.}~\bibnamefont {Vanderbilt}},\ }\bibfield
   {title} {\enquote {\bibinfo {title} {Controllable quantum point junction on
  the surface of an antiferromagnetic topological insulator}}, }\href {\doibase
  10.1038/s41467-021-24276-5} {\bibfield  {journal} {\bibinfo  {journal} {Nat.
  Commun.}\ }\textbf {\bibinfo {volume} {12}},\ \bibinfo {pages} {3998}
  (\bibinfo {year} {2021})}\BibitemShut {NoStop}%
\bibitem [{\citenamefont {Gu}\ \emph {et~al.}(2021)\citenamefont {Gu},
  \citenamefont {Li}, \citenamefont {Sun}, \citenamefont {Zhao}, \citenamefont
  {Liu}, \citenamefont {Liu}, \citenamefont {Lu},\ and\ \citenamefont
  {Liu}}]{Gu21NCSpectral}%
  \BibitemOpen
  \bibfield  {author} {\bibinfo {author} {\bibfnamefont {M.}~\bibnamefont
  {Gu}}, \bibinfo {author} {\bibfnamefont {J.}~\bibnamefont {Li}}, \bibinfo
  {author} {\bibfnamefont {H.}~\bibnamefont {Sun}}, \bibinfo {author}
  {\bibfnamefont {Y.}~\bibnamefont {Zhao}}, \bibinfo {author} {\bibfnamefont
  {C.}~\bibnamefont {Liu}}, \bibinfo {author} {\bibfnamefont {J.}~\bibnamefont
  {Liu}}, \bibinfo {author} {\bibfnamefont {H.}~\bibnamefont {Lu}}, \ and\
  \bibinfo {author} {\bibfnamefont {Q.}~\bibnamefont {Liu}},\ }\bibfield
  {title} {\enquote {\bibinfo {title} {Spectral signatures of the surface
  anomalous {{Hall}} effect in magnetic axion insulators}}, }\href {\doibase
  10.1038/s41467-021-23844-z} {\bibfield  {journal} {\bibinfo  {journal} {Nat.
  Commun.}\ }\textbf {\bibinfo {volume} {12}},\ \bibinfo {pages} {3524}
  (\bibinfo {year} {2021})}\BibitemShut {NoStop}%
\bibitem [{\citenamefont {Li}\ \emph {et~al.}(2021)\citenamefont {Li},
  \citenamefont {Chen}, \citenamefont {Jiang},\ and\ \citenamefont
  {Xie}}]{Li21PRLCoexistence}%
  \BibitemOpen
  \bibfield  {author} {\bibinfo {author} {\bibfnamefont {H.}~\bibnamefont
  {Li}}, \bibinfo {author} {\bibfnamefont {C.-Z.}\ \bibnamefont {Chen}},
  \bibinfo {author} {\bibfnamefont {H.}~\bibnamefont {Jiang}}, \ and\ \bibinfo
  {author} {\bibfnamefont {X.~C.}\ \bibnamefont {Xie}},\ }\bibfield  {title}
  {\enquote {\bibinfo {title} {Coexistence of {{Quantum Hall}} and {{Quantum
  Anomalous Hall Phases}} in {{Disordered MnBi}}$_2${{Te}}$_4$}}, }\href
  {\doibase 10.1103/PhysRevLett.127.236402} {\bibfield  {journal} {\bibinfo
  {journal} {Phys. Rev. Lett.}\ }\textbf {\bibinfo {volume} {127}},\ \bibinfo
  {pages} {236402} (\bibinfo {year} {2021})}\BibitemShut {NoStop}%
\bibitem [{\citenamefont {Chen}\ \emph
  {et~al.}(2021{\natexlab{a}})\citenamefont {Chen}, \citenamefont {Li},
  \citenamefont {Sun}, \citenamefont {Liu}, \citenamefont {Zhao}, \citenamefont
  {Lu},\ and\ \citenamefont {Xie}}]{Chen21PRBUsing}%
  \BibitemOpen
  \bibfield  {author} {\bibinfo {author} {\bibfnamefont {R.}~\bibnamefont
  {Chen}}, \bibinfo {author} {\bibfnamefont {S.}~\bibnamefont {Li}}, \bibinfo
  {author} {\bibfnamefont {H.-P.}\ \bibnamefont {Sun}}, \bibinfo {author}
  {\bibfnamefont {Q.}~\bibnamefont {Liu}}, \bibinfo {author} {\bibfnamefont
  {Y.}~\bibnamefont {Zhao}}, \bibinfo {author} {\bibfnamefont {H.-Z.}\
  \bibnamefont {Lu}}, \ and\ \bibinfo {author} {\bibfnamefont {X.~C.}\
  \bibnamefont {Xie}},\ }\bibfield  {title} {\enquote {\bibinfo {title} {Using
  nonlocal surface transport to identify the axion insulator}}, }\href
  {\doibase 10.1103/PhysRevB.103.L241409} {\bibfield  {journal} {\bibinfo
  {journal} {Phys. Rev. B}\ }\textbf {\bibinfo {volume} {103}},\ \bibinfo
  {pages} {L241409} (\bibinfo {year} {2021}{\natexlab{a}})}\BibitemShut
  {NoStop}%
\bibitem [{\citenamefont {Chen}\ \emph
  {et~al.}(2021{\natexlab{b}})\citenamefont {Chen}, \citenamefont {Zhao},
  \citenamefont {Yao}, \citenamefont {Zhang},\ and\ \citenamefont
  {Liu}}]{Chen21PRBKoopmans}%
  \BibitemOpen
  \bibfield  {author} {\bibinfo {author} {\bibfnamefont {W.}~\bibnamefont
  {Chen}}, \bibinfo {author} {\bibfnamefont {Y.}~\bibnamefont {Zhao}}, \bibinfo
  {author} {\bibfnamefont {Q.}~\bibnamefont {Yao}}, \bibinfo {author}
  {\bibfnamefont {J.}~\bibnamefont {Zhang}}, \ and\ \bibinfo {author}
  {\bibfnamefont {Q.}~\bibnamefont {Liu}},\ }\bibfield  {title} {\enquote
  {\bibinfo {title} {Koopmans' theorem as the mechanism of nearly gapless
  surface states in self-doped magnetic topological insulators}}, }\href
  {\doibase 10.1103/PhysRevB.103.L201102} {\bibfield  {journal} {\bibinfo
  {journal} {Phys. Rev. B}\ }\textbf {\bibinfo {volume} {103}},\ \bibinfo
  {pages} {L201102} (\bibinfo {year} {2021}{\natexlab{b}})}\BibitemShut
  {NoStop}%
\bibitem [{\citenamefont {Du}\ \emph {et~al.}(2020)\citenamefont {Du},
  \citenamefont {Tang}, \citenamefont {Li}, \citenamefont {Lin}, \citenamefont
  {Xu}, \citenamefont {Duan},\ and\ \citenamefont {Rubio}}]{Du20PRRBerry}%
  \BibitemOpen
  \bibfield  {author} {\bibinfo {author} {\bibfnamefont {S.}~\bibnamefont
  {Du}}, \bibinfo {author} {\bibfnamefont {P.}~\bibnamefont {Tang}}, \bibinfo
  {author} {\bibfnamefont {J.}~\bibnamefont {Li}}, \bibinfo {author}
  {\bibfnamefont {Z.}~\bibnamefont {Lin}}, \bibinfo {author} {\bibfnamefont
  {Y.}~\bibnamefont {Xu}}, \bibinfo {author} {\bibfnamefont {W.}~\bibnamefont
  {Duan}}, \ and\ \bibinfo {author} {\bibfnamefont {A.}~\bibnamefont {Rubio}},\
  }\bibfield  {title} {\enquote {\bibinfo {title} {Berry curvature engineering
  by gating two-dimensional antiferromagnets}}, }\href {\doibase
  10.1103/PhysRevResearch.2.022025} {\bibfield  {journal} {\bibinfo  {journal}
  {Phys. Rev. Research}\ }\textbf {\bibinfo {volume} {2}},\ \bibinfo {pages}
  {022025} (\bibinfo {year} {2020})}\BibitemShut {NoStop}%
\bibitem [{\citenamefont {Liu}\ and\ \citenamefont
  {Wang}(2020)}]{Liu20PRBAnisotropic}%
  \BibitemOpen
  \bibfield  {author} {\bibinfo {author} {\bibfnamefont {Z.}~\bibnamefont
  {Liu}}\ and\ \bibinfo {author} {\bibfnamefont {J.}~\bibnamefont {Wang}},\
  }\bibfield  {title} {\enquote {\bibinfo {title} {Anisotropic topological
  magnetoelectric effect in axion insulators}}, }\href {\doibase
  10.1103/PhysRevB.101.205130} {\bibfield  {journal} {\bibinfo  {journal}
  {Phys. Rev. B}\ }\textbf {\bibinfo {volume} {101}},\ \bibinfo {pages}
  {205130} (\bibinfo {year} {2020})}\BibitemShut {NoStop}%
\bibitem [{\citenamefont {Liu}\ \emph {et~al.}(2021{\natexlab{a}})\citenamefont
  {Liu}, \citenamefont {Qian}, \citenamefont {Jiang},\ and\ \citenamefont
  {Wang}}]{Liu21Dissipative}%
  \BibitemOpen
  \bibfield  {author} {\bibinfo {author} {\bibfnamefont {Z.}~\bibnamefont
  {Liu}}, \bibinfo {author} {\bibfnamefont {D.}~\bibnamefont {Qian}}, \bibinfo
  {author} {\bibfnamefont {Y.}~\bibnamefont {Jiang}}, \ and\ \bibinfo {author}
  {\bibfnamefont {J.}~\bibnamefont {Wang}},\ }\href {\doibase
  10.48550/arXiv.2109.06178} {\enquote {\bibinfo {title} {Dissipative {{Edge
  Transport}} in {{Disordered Axion Insulator Films}}}}, } (\bibinfo {year}
  {2021}{\natexlab{a}}),\ \Eprint {http://arxiv.org/abs/2109.06178}
  {arXiv:2109.06178} \BibitemShut {NoStop}%
\bibitem [{\citenamefont {Cai}\ \emph {et~al.}(2022)\citenamefont {Cai},
  \citenamefont {Ovchinnikov}, \citenamefont {Fei}, \citenamefont {He},
  \citenamefont {Song}, \citenamefont {Lin}, \citenamefont {Wang},
  \citenamefont {Cobden}, \citenamefont {Chu}, \citenamefont {Cui},
  \citenamefont {Chang}, \citenamefont {Xiao}, \citenamefont {Yan},\ and\
  \citenamefont {Xu}}]{Cai22NCElectric}%
  \BibitemOpen
  \bibfield  {author} {\bibinfo {author} {\bibfnamefont {J.}~\bibnamefont
  {Cai}},  \emph {et~al.},\ }\bibfield  {title} {\enquote {\bibinfo {title}
  {Electric control of a canted-antiferromagnetic {{Chern}} insulator}}, }\href
  {\doibase 10.1038/s41467-022-29259-8} {\bibfield  {journal} {\bibinfo
  {journal} {Nat. Commun.}\ }\textbf {\bibinfo {volume} {13}},\ \bibinfo
  {pages} {1668} (\bibinfo {year} {2022})}\BibitemShut {NoStop}%
\bibitem [{\citenamefont {Gao}\ \emph {et~al.}(2021)\citenamefont {Gao},
  \citenamefont {Liu}, \citenamefont {Hu}, \citenamefont {Qiu}, \citenamefont
  {Tzschaschel}, \citenamefont {Ghosh}, \citenamefont {Ho}, \citenamefont
  {B{\'e}rub{\'e}}, \citenamefont {Chen}, \citenamefont {Sun}, \citenamefont
  {Zhang}, \citenamefont {Zhang}, \citenamefont {Wang}, \citenamefont {Wang},
  \citenamefont {Huang}, \citenamefont {Felser}, \citenamefont {Agarwal},
  \citenamefont {Ding}, \citenamefont {Tien}, \citenamefont {Akey},
  \citenamefont {Gardener}, \citenamefont {Singh}, \citenamefont {Watanabe},
  \citenamefont {Taniguchi}, \citenamefont {Burch}, \citenamefont {Bell},
  \citenamefont {Zhou}, \citenamefont {Gao}, \citenamefont {Lu}, \citenamefont
  {Bansil}, \citenamefont {Lin}, \citenamefont {Chang}, \citenamefont {Fu},
  \citenamefont {Ma}, \citenamefont {Ni},\ and\ \citenamefont
  {Xu}}]{Gao21NLayer}%
  \BibitemOpen
  \bibfield  {author} {\bibinfo {author} {\bibfnamefont {A.}~\bibnamefont
  {Gao}},  \emph {et~al.},\ }\bibfield  {title} {\enquote {\bibinfo {title}
  {Layer {{Hall}} effect in a {{2D}} topological axion antiferromagnet}},
  }\href {\doibase 10.1038/s41586-021-03679-w} {\bibfield  {journal} {\bibinfo
  {journal} {Nature}\ }\textbf {\bibinfo {volume} {595}},\ \bibinfo {pages}
  {521} (\bibinfo {year} {2021})}\BibitemShut {NoStop}%
\bibitem [{\citenamefont {Liu}\ \emph {et~al.}(2021{\natexlab{b}})\citenamefont
  {Liu}, \citenamefont {Wang}, \citenamefont {Yang}, \citenamefont {Mao},
  \citenamefont {Li}, \citenamefont {Li}, \citenamefont {Li}, \citenamefont
  {Zhu}, \citenamefont {Wang}, \citenamefont {Li}, \citenamefont {Wu},
  \citenamefont {Xu}, \citenamefont {Zhang},\ and\ \citenamefont
  {Wang}}]{Liu21NCMagneticfieldinduced}%
  \BibitemOpen
  \bibfield  {author} {\bibinfo {author} {\bibfnamefont {C.}~\bibnamefont
  {Liu}},  \emph {et~al.},\ }\bibfield  {title} {\enquote {\bibinfo {title}
  {Magnetic-field-induced robust zero {{Hall}} plateau state in
  {{MnBi}}$_2${{Te}}$_4$ {{Chern}} insulator}}, }\href {\doibase
  10.1038/s41467-021-25002-x} {\bibfield  {journal} {\bibinfo  {journal} {Nat.
  Commun.}\ }\textbf {\bibinfo {volume} {12}},\ \bibinfo {pages} {4647}
  (\bibinfo {year} {2021}{\natexlab{b}})}\BibitemShut {NoStop}%
\bibitem [{\citenamefont {Chen}\ \emph
  {et~al.}(2019{\natexlab{a}})\citenamefont {Chen}, \citenamefont {Fei},
  \citenamefont {Zhang}, \citenamefont {Zhang}, \citenamefont {Liu},
  \citenamefont {Zhang}, \citenamefont {Wang}, \citenamefont {Wei},
  \citenamefont {Zhang}, \citenamefont {Zuo}, \citenamefont {Guo},
  \citenamefont {Liu}, \citenamefont {Wang}, \citenamefont {Wu}, \citenamefont
  {Zong}, \citenamefont {Xie}, \citenamefont {Chen}, \citenamefont {Sun},
  \citenamefont {Wang}, \citenamefont {Zhang}, \citenamefont {Zhang},
  \citenamefont {Wang}, \citenamefont {Song}, \citenamefont {Zhang},
  \citenamefont {Shen},\ and\ \citenamefont {Wang}}]{Chen19NCIntrinsic}%
  \BibitemOpen
  \bibfield  {author} {\bibinfo {author} {\bibfnamefont {B.}~\bibnamefont
  {Chen}},  \emph {et~al.},\ }\bibfield  {title} {\enquote {\bibinfo {title}
  {Intrinsic magnetic topological insulator phases in the {{Sb}} doped
  {{MnBi}}$_2${{Te}}$_4$ bulks and thin flakes}}, }\href
  {https://www.nature.com/articles/s41467-019-12485-y} {\bibfield  {journal}
  {\bibinfo  {journal} {Nat. Commun.}\ }\textbf {\bibinfo {volume} {10}},\
  \bibinfo {pages} {4469} (\bibinfo {year} {2019}{\natexlab{a}})}\BibitemShut
  {NoStop}%
\bibitem [{\citenamefont {Hao}\ \emph {et~al.}(2019)\citenamefont {Hao},
  \citenamefont {Liu}, \citenamefont {Feng}, \citenamefont {Ma}, \citenamefont
  {Schwier}, \citenamefont {Arita}, \citenamefont {Kumar}, \citenamefont {Hu},
  \citenamefont {Lu}, \citenamefont {Zeng}, \citenamefont {Wang}, \citenamefont
  {Hao}, \citenamefont {Sun}, \citenamefont {Zhang}, \citenamefont {Mei},
  \citenamefont {Ni}, \citenamefont {Wu}, \citenamefont {Shimada},
  \citenamefont {Chen}, \citenamefont {Liu},\ and\ \citenamefont
  {Liu}}]{Hao19PRXGapless}%
  \BibitemOpen
  \bibfield  {author} {\bibinfo {author} {\bibfnamefont {Y.-J.}\ \bibnamefont
  {Hao}},  \emph {et~al.},\ }\bibfield  {title} {\enquote {\bibinfo {title}
  {Gapless {{Surface Dirac Cone}} in {{Antiferromagnetic Topological Insulator
  MnBi}}$_2${{Te}}$_4$}}, }\href
  {https://link.aps.org/doi/10.1103/PhysRevX.9.041038} {\bibfield  {journal}
  {\bibinfo  {journal} {Phys. Rev. X}\ }\textbf {\bibinfo {volume} {9}},\
  \bibinfo {pages} {041038} (\bibinfo {year} {2019})}\BibitemShut {NoStop}%
\bibitem [{\citenamefont {Li}\ \emph {et~al.}(2019{\natexlab{b}})\citenamefont
  {Li}, \citenamefont {Gao}, \citenamefont {Duan}, \citenamefont {Xu},
  \citenamefont {Zhu}, \citenamefont {Tian}, \citenamefont {Gao}, \citenamefont
  {Fan}, \citenamefont {Rao}, \citenamefont {Huang}, \citenamefont {Li},
  \citenamefont {Yan}, \citenamefont {Liu}, \citenamefont {Liu}, \citenamefont
  {Huang}, \citenamefont {Li}, \citenamefont {Liu}, \citenamefont {Zhang},
  \citenamefont {Zhang}, \citenamefont {Kondo}, \citenamefont {Shin},
  \citenamefont {Lei}, \citenamefont {Shi}, \citenamefont {Zhang},
  \citenamefont {Weng}, \citenamefont {Qian},\ and\ \citenamefont
  {Ding}}]{Li19PRXDirac}%
  \BibitemOpen
  \bibfield  {author} {\bibinfo {author} {\bibfnamefont {H.}~\bibnamefont
  {Li}},  \emph {et~al.},\ }\bibfield  {title} {\enquote {\bibinfo {title}
  {Dirac {{Surface States}} in {{Intrinsic Magnetic Topological Insulators
  EuSn}}$_2${{As}}$_2$ and {{MnBi}}$_{2n}${{Te}}$_{3n+1}$}}, }\href
  {https://link.aps.org/doi/10.1103/PhysRevX.9.041039} {\bibfield  {journal}
  {\bibinfo  {journal} {Phys. Rev. X}\ }\textbf {\bibinfo {volume} {9}},\
  \bibinfo {pages} {041039} (\bibinfo {year} {2019}{\natexlab{b}})}\BibitemShut
  {NoStop}%
\bibitem [{\citenamefont {Chen}\ \emph
  {et~al.}(2019{\natexlab{b}})\citenamefont {Chen}, \citenamefont {Xu},
  \citenamefont {Li}, \citenamefont {Li}, \citenamefont {Wang}, \citenamefont
  {Zhang}, \citenamefont {Li}, \citenamefont {Wu}, \citenamefont {Liang},
  \citenamefont {Chen}, \citenamefont {Jung}, \citenamefont {Cacho},
  \citenamefont {Mao}, \citenamefont {Liu}, \citenamefont {Wang}, \citenamefont
  {Guo}, \citenamefont {Xu}, \citenamefont {Liu}, \citenamefont {Yang},\ and\
  \citenamefont {Chen}}]{Chen19PRXTopological}%
  \BibitemOpen
  \bibfield  {author} {\bibinfo {author} {\bibfnamefont {Y.~J.}\ \bibnamefont
  {Chen}},  \emph {et~al.},\ }\bibfield  {title} {\enquote {\bibinfo {title}
  {Topological {{Electronic Structure}} and {{Its Temperature Evolution}} in
  {{Antiferromagnetic Topological Insulator MnBi}}$_2${{Te}}$_4$}}, }\href
  {https://link.aps.org/doi/10.1103/PhysRevX.9.041040} {\bibfield  {journal}
  {\bibinfo  {journal} {Phys. Rev. X}\ }\textbf {\bibinfo {volume} {9}},\
  \bibinfo {pages} {041040} (\bibinfo {year} {2019}{\natexlab{b}})}\BibitemShut
  {NoStop}%
\bibitem [{\citenamefont {Swatek}\ \emph {et~al.}(2020)\citenamefont {Swatek},
  \citenamefont {Wu}, \citenamefont {Wang}, \citenamefont {Lee}, \citenamefont
  {Schrunk}, \citenamefont {Yan},\ and\ \citenamefont
  {Kaminski}}]{Swatek20PRBGapless}%
  \BibitemOpen
  \bibfield  {author} {\bibinfo {author} {\bibfnamefont {P.}~\bibnamefont
  {Swatek}}, \bibinfo {author} {\bibfnamefont {Y.}~\bibnamefont {Wu}}, \bibinfo
  {author} {\bibfnamefont {L.-L.}\ \bibnamefont {Wang}}, \bibinfo {author}
  {\bibfnamefont {K.}~\bibnamefont {Lee}}, \bibinfo {author} {\bibfnamefont
  {B.}~\bibnamefont {Schrunk}}, \bibinfo {author} {\bibfnamefont
  {J.}~\bibnamefont {Yan}}, \ and\ \bibinfo {author} {\bibfnamefont
  {A.}~\bibnamefont {Kaminski}},\ }\bibfield  {title} {\enquote {\bibinfo
  {title} {Gapless {{Dirac}} surface states in the antiferromagnetic
  topological insulator {{MnBi}}$_2${{Te}}$_4$}}, }\href
  {https://link.aps.org/doi/10.1103/PhysRevB.101.161109} {\bibfield  {journal}
  {\bibinfo  {journal} {Phys. Rev. B}\ }\textbf {\bibinfo {volume} {101}},\
  \bibinfo {pages} {161109} (\bibinfo {year} {2020})}\BibitemShut {NoStop}%
\bibitem [{\citenamefont {Wu}\ \emph {et~al.}(2020)\citenamefont {Wu},
  \citenamefont {Li}, \citenamefont {Ma}, \citenamefont {Zhang}, \citenamefont
  {Liu}, \citenamefont {Zhou}, \citenamefont {Shao}, \citenamefont {Wang},
  \citenamefont {Hao}, \citenamefont {Feng}, \citenamefont {Schwier},
  \citenamefont {Kumar}, \citenamefont {Sun}, \citenamefont {Liu},
  \citenamefont {Shimada}, \citenamefont {Miyamoto}, \citenamefont {Okuda},
  \citenamefont {Wang}, \citenamefont {Xie}, \citenamefont {Chen},
  \citenamefont {Liu}, \citenamefont {Liu},\ and\ \citenamefont
  {Zhao}}]{Wu20PRXDistinct}%
  \BibitemOpen
  \bibfield  {author} {\bibinfo {author} {\bibfnamefont {X.}~\bibnamefont
  {Wu}},  \emph {et~al.},\ }\bibfield  {title} {\enquote {\bibinfo {title}
  {Distinct {{Topological Surface States}} on the {{Two Terminations}} of
  {{MnBi}}$_4${{Te}}$_7$}}, }\href {\doibase 10.1103/PhysRevX.10.031013}
  {\bibfield  {journal} {\bibinfo  {journal} {Phys. Rev. X}\ }\textbf {\bibinfo
  {volume} {10}},\ \bibinfo {pages} {031013} (\bibinfo {year}
  {2020})}\BibitemShut {NoStop}%
\bibitem [{\citenamefont {Lu}\ \emph {et~al.}(2021)\citenamefont {Lu},
  \citenamefont {Sun}, \citenamefont {Kumar}, \citenamefont {Wang},
  \citenamefont {Gu}, \citenamefont {Zeng}, \citenamefont {Hao}, \citenamefont
  {Li}, \citenamefont {Shao}, \citenamefont {Ma}, \citenamefont {Hao},
  \citenamefont {Zhang}, \citenamefont {Mansuer}, \citenamefont {Mei},
  \citenamefont {Zhao}, \citenamefont {Liu}, \citenamefont {Deng},
  \citenamefont {Huang}, \citenamefont {Shen}, \citenamefont {Shimada},
  \citenamefont {Schwier}, \citenamefont {Liu}, \citenamefont {Liu},\ and\
  \citenamefont {Chen}}]{Lu21PRXHalfMagnetic}%
  \BibitemOpen
  \bibfield  {author} {\bibinfo {author} {\bibfnamefont {R.}~\bibnamefont
  {Lu}},  \emph {et~al.},\ }\bibfield  {title} {\enquote {\bibinfo {title}
  {Half-{{Magnetic Topological Insulator}} with {{Magnetization-Induced Dirac
  Gap}} at a {{Selected Surface}}}}, }\href {\doibase
  10.1103/PhysRevX.11.011039} {\bibfield  {journal} {\bibinfo  {journal} {Phys.
  Rev. X}\ }\textbf {\bibinfo {volume} {11}},\ \bibinfo {pages} {011039}
  (\bibinfo {year} {2021})}\BibitemShut {NoStop}%
\bibitem [{\citenamefont {Vidal}\ \emph {et~al.}(2021)\citenamefont {Vidal},
  \citenamefont {Bentmann}, \citenamefont {Facio}, \citenamefont {Heider},
  \citenamefont {Kagerer}, \citenamefont {Fornari}, \citenamefont {Peixoto},
  \citenamefont {Figgemeier}, \citenamefont {Jung}, \citenamefont {Cacho},
  \citenamefont {B{\"u}chner}, \citenamefont {{van den Brink}}, \citenamefont
  {Schneider}, \citenamefont {Plucinski}, \citenamefont {Schwier},
  \citenamefont {Shimada}, \citenamefont {Richter}, \citenamefont {Isaeva},\
  and\ \citenamefont {Reinert}}]{Vidal21PRLOrbital}%
  \BibitemOpen
  \bibfield  {author} {\bibinfo {author} {\bibfnamefont {R.~C.}\ \bibnamefont
  {Vidal}},  \emph {et~al.},\ }\bibfield  {title} {\enquote {\bibinfo {title}
  {Orbital {{Complexity}} in {{Intrinsic Magnetic Topological Insulators
  MnBi}}$_4${{Te}}$_7$ and {{MnBi}}$_6${{Te}}$_10$}}, }\href {\doibase
  10.1103/PhysRevLett.126.176403} {\bibfield  {journal} {\bibinfo  {journal}
  {Phys. Rev. Lett.}\ }\textbf {\bibinfo {volume} {126}},\ \bibinfo {pages}
  {176403} (\bibinfo {year} {2021})}\BibitemShut {NoStop}%
\bibitem [{\citenamefont {Lee}\ \emph {et~al.}(2021)\citenamefont {Lee},
  \citenamefont {Graf}, \citenamefont {Min}, \citenamefont {Zhu}, \citenamefont
  {Yi}, \citenamefont {Ciocys}, \citenamefont {Wang}, \citenamefont {Choi},
  \citenamefont {Basnet}, \citenamefont {Fereidouni}, \citenamefont {Wegner},
  \citenamefont {Zhao}, \citenamefont {Verlinde}, \citenamefont {He},
  \citenamefont {Redwing}, \citenamefont {Gopalan}, \citenamefont {Churchill},
  \citenamefont {Lanzara}, \citenamefont {Samarth}, \citenamefont {Chang},
  \citenamefont {Hu},\ and\ \citenamefont {Mao}}]{Lee21PRXEvidence}%
  \BibitemOpen
  \bibfield  {author} {\bibinfo {author} {\bibfnamefont {S.~H.}\ \bibnamefont
  {Lee}},  \emph {et~al.},\ }\bibfield  {title} {\enquote {\bibinfo {title}
  {Evidence for a {{Magnetic-Field-Induced Ideal Type-II Weyl State}} in
  {{Antiferromagnetic Topological Insulator
  Mn}}({{Bi}}$_{1-x}${{Sb}}$_x$)$_2${{Te}}$_4$}}, }\href {\doibase
  10.1103/PhysRevX.11.031032} {\bibfield  {journal} {\bibinfo  {journal} {Phys.
  Rev. X}\ }\textbf {\bibinfo {volume} {11}},\ \bibinfo {pages} {031032}
  (\bibinfo {year} {2021})}\BibitemShut {NoStop}%
\bibitem [{\citenamefont {Deng}\ \emph {et~al.}(2020)\citenamefont {Deng},
  \citenamefont {Yu}, \citenamefont {Shi}, \citenamefont {Guo}, \citenamefont
  {Xu}, \citenamefont {Wang}, \citenamefont {Chen},\ and\ \citenamefont
  {Zhang}}]{Deng20SQuantum}%
  \BibitemOpen
  \bibfield  {author} {\bibinfo {author} {\bibfnamefont {Y.}~\bibnamefont
  {Deng}}, \bibinfo {author} {\bibfnamefont {Y.}~\bibnamefont {Yu}}, \bibinfo
  {author} {\bibfnamefont {M.~Z.}\ \bibnamefont {Shi}}, \bibinfo {author}
  {\bibfnamefont {Z.}~\bibnamefont {Guo}}, \bibinfo {author} {\bibfnamefont
  {Z.}~\bibnamefont {Xu}}, \bibinfo {author} {\bibfnamefont {J.}~\bibnamefont
  {Wang}}, \bibinfo {author} {\bibfnamefont {X.~H.}\ \bibnamefont {Chen}}, \
  and\ \bibinfo {author} {\bibfnamefont {Y.}~\bibnamefont {Zhang}},\ }\bibfield
   {title} {\enquote {\bibinfo {title} {Quantum anomalous {{Hall}} effect in
  intrinsic magnetic topological insulator {{MnBi}}$_2${{Te}}$_4$}}, }\href
  {https://science.sciencemag.org/content/367/6480/895} {\bibfield  {journal}
  {\bibinfo  {journal} {Science}\ }\textbf {\bibinfo {volume} {367}},\ \bibinfo
  {pages} {895} (\bibinfo {year} {2020})}\BibitemShut {NoStop}%
\bibitem [{\citenamefont {Liu}\ \emph {et~al.}(2020)\citenamefont {Liu},
  \citenamefont {Wang}, \citenamefont {Li}, \citenamefont {Wu}, \citenamefont
  {Li}, \citenamefont {Li}, \citenamefont {He}, \citenamefont {Xu},
  \citenamefont {Zhang},\ and\ \citenamefont {Wang}}]{Liu20NMRobust}%
  \BibitemOpen
  \bibfield  {author} {\bibinfo {author} {\bibfnamefont {C.}~\bibnamefont
  {Liu}}, \bibinfo {author} {\bibfnamefont {Y.}~\bibnamefont {Wang}}, \bibinfo
  {author} {\bibfnamefont {H.}~\bibnamefont {Li}}, \bibinfo {author}
  {\bibfnamefont {Y.}~\bibnamefont {Wu}}, \bibinfo {author} {\bibfnamefont
  {Y.}~\bibnamefont {Li}}, \bibinfo {author} {\bibfnamefont {J.}~\bibnamefont
  {Li}}, \bibinfo {author} {\bibfnamefont {K.}~\bibnamefont {He}}, \bibinfo
  {author} {\bibfnamefont {Y.}~\bibnamefont {Xu}}, \bibinfo {author}
  {\bibfnamefont {J.}~\bibnamefont {Zhang}}, \ and\ \bibinfo {author}
  {\bibfnamefont {Y.}~\bibnamefont {Wang}},\ }\bibfield  {title} {\enquote
  {\bibinfo {title} {Robust axion insulator and {{Chern}} insulator phases in a
  two-dimensional antiferromagnetic topological insulator}}, }\href
  {https://www.nature.com/articles/s41563-019-0573-3} {\bibfield  {journal}
  {\bibinfo  {journal} {Nat. Mater.}\ }\textbf {\bibinfo {volume} {19}},\
  \bibinfo {pages} {522} (\bibinfo {year} {2020})}\BibitemShut {NoStop}%
\bibitem [{\citenamefont {Deng}\ \emph {et~al.}(2021)\citenamefont {Deng},
  \citenamefont {Chen}, \citenamefont {Wo{\l}o{\'s}}, \citenamefont
  {Konczykowski}, \citenamefont {Sobczak}, \citenamefont {Sitnicka},
  \citenamefont {Fedorchenko}, \citenamefont {Borysiuk}, \citenamefont
  {Heider}, \citenamefont {Pluci{\'n}ski}, \citenamefont {Park}, \citenamefont
  {Georgescu}, \citenamefont {Cano},\ and\ \citenamefont
  {{Krusin-Elbaum}}}]{Deng21NPHightemperature}%
  \BibitemOpen
  \bibfield  {author} {\bibinfo {author} {\bibfnamefont {H.}~\bibnamefont
  {Deng}},  \emph {et~al.},\ }\bibfield  {title} {\enquote {\bibinfo {title}
  {High-temperature quantum anomalous {{Hall}} regime in a
  {{MnBi}}$_2${{Te}}$_4$/{{Bi}}$_2${{Te}}$_3$ superlattice}}, }\href {\doibase
  10.1038/s41567-020-0998-2} {\bibfield  {journal} {\bibinfo  {journal} {Nat.
  Phys.}\ }\textbf {\bibinfo {volume} {17}},\ \bibinfo {pages} {36} (\bibinfo
  {year} {2021})}\BibitemShut {NoStop}%
\bibitem [{\citenamefont {Ge}\ \emph {et~al.}(2020)\citenamefont {Ge},
  \citenamefont {Liu}, \citenamefont {Li}, \citenamefont {Li}, \citenamefont
  {Luo}, \citenamefont {Wu}, \citenamefont {Xu},\ and\ \citenamefont
  {Wang}}]{Ge20NSRHighChernnumber}%
  \BibitemOpen
  \bibfield  {author} {\bibinfo {author} {\bibfnamefont {J.}~\bibnamefont
  {Ge}}, \bibinfo {author} {\bibfnamefont {Y.}~\bibnamefont {Liu}}, \bibinfo
  {author} {\bibfnamefont {J.}~\bibnamefont {Li}}, \bibinfo {author}
  {\bibfnamefont {H.}~\bibnamefont {Li}}, \bibinfo {author} {\bibfnamefont
  {T.}~\bibnamefont {Luo}}, \bibinfo {author} {\bibfnamefont {Y.}~\bibnamefont
  {Wu}}, \bibinfo {author} {\bibfnamefont {Y.}~\bibnamefont {Xu}}, \ and\
  \bibinfo {author} {\bibfnamefont {J.}~\bibnamefont {Wang}},\ }\bibfield
  {title} {\enquote {\bibinfo {title} {High-{{Chern-number}} and
  high-temperature quantum {{Hall}} effect without {{Landau}} levels}}, }\href
  {https://doi.org/10.1093/nsr/nwaa089} {\bibfield  {journal} {\bibinfo
  {journal} {Natl. Sci. Rev.}\ }\textbf {\bibinfo {volume} {7}},\ \bibinfo
  {pages} {1280} (\bibinfo {year} {2020})}\BibitemShut {NoStop}%
\bibitem [{\citenamefont {Ovchinnikov}\ \emph {et~al.}(2021)\citenamefont
  {Ovchinnikov}, \citenamefont {Huang}, \citenamefont {Lin}, \citenamefont
  {Fei}, \citenamefont {Cai}, \citenamefont {Song}, \citenamefont {He},
  \citenamefont {Jiang}, \citenamefont {Wang}, \citenamefont {Li},
  \citenamefont {Wang}, \citenamefont {Wu}, \citenamefont {Xiao}, \citenamefont
  {Chu}, \citenamefont {Yan}, \citenamefont {Chang}, \citenamefont {Cui},\ and\
  \citenamefont {Xu}}]{Ovchinnikov21NLIntertwined}%
  \BibitemOpen
  \bibfield  {author} {\bibinfo {author} {\bibfnamefont {D.}~\bibnamefont
  {Ovchinnikov}},  \emph {et~al.},\ }\bibfield  {title} {\enquote {\bibinfo
  {title} {Intertwined {{Topological}} and {{Magnetic Orders}} in {{Atomically
  Thin Chern Insulator MnBi}}$_2${{Te}}$_4$}}, }\href {\doibase
  10.1021/acs.nanolett.0c05117} {\bibfield  {journal} {\bibinfo  {journal}
  {Nano Lett.}\ }\textbf {\bibinfo {volume} {21}},\ \bibinfo {pages} {2544}
  (\bibinfo {year} {2021})}\BibitemShut {NoStop}%
\bibitem [{\citenamefont {Ge}\ \emph {et~al.}(2022)\citenamefont {Ge},
  \citenamefont {Liu}, \citenamefont {Wang}, \citenamefont {Xu}, \citenamefont
  {Li}, \citenamefont {Li}, \citenamefont {Yan}, \citenamefont {Wu},
  \citenamefont {Xu},\ and\ \citenamefont {Wang}}]{Ge22PRBMagnetizationtuned}%
  \BibitemOpen
  \bibfield  {author} {\bibinfo {author} {\bibfnamefont {J.}~\bibnamefont
  {Ge}}, \bibinfo {author} {\bibfnamefont {Y.}~\bibnamefont {Liu}}, \bibinfo
  {author} {\bibfnamefont {P.}~\bibnamefont {Wang}}, \bibinfo {author}
  {\bibfnamefont {Z.}~\bibnamefont {Xu}}, \bibinfo {author} {\bibfnamefont
  {J.}~\bibnamefont {Li}}, \bibinfo {author} {\bibfnamefont {H.}~\bibnamefont
  {Li}}, \bibinfo {author} {\bibfnamefont {Z.}~\bibnamefont {Yan}}, \bibinfo
  {author} {\bibfnamefont {Y.}~\bibnamefont {Wu}}, \bibinfo {author}
  {\bibfnamefont {Y.}~\bibnamefont {Xu}}, \ and\ \bibinfo {author}
  {\bibfnamefont {J.}~\bibnamefont {Wang}},\ }\bibfield  {title} {\enquote
  {\bibinfo {title} {Magnetization-tuned topological quantum phase transition
  in {{MnBi}}$_2${{Te}}$_4$ devices}}, }\href {\doibase
  10.1103/PhysRevB.105.L201404} {\bibfield  {journal} {\bibinfo  {journal}
  {Phys. Rev. B}\ }\textbf {\bibinfo {volume} {105}},\ \bibinfo {pages}
  {L201404} (\bibinfo {year} {2022})}\BibitemShut {NoStop}%
\bibitem [{\citenamefont {Liang}\ \emph {et~al.}(2022)\citenamefont {Liang},
  \citenamefont {Chen}, \citenamefont {Zheng}, \citenamefont {Xia},
  \citenamefont {Huang}, \citenamefont {Wei}, \citenamefont {Yang},
  \citenamefont {Chen}, \citenamefont {Zhang}, \citenamefont {Xu},
  \citenamefont {Wang}, \citenamefont {Guo}, \citenamefont {Yang},
  \citenamefont {Liu},\ and\ \citenamefont {Chen}}]{Liang22NLApproaching}%
  \BibitemOpen
  \bibfield  {author} {\bibinfo {author} {\bibfnamefont {A.}~\bibnamefont
  {Liang}},  \emph {et~al.},\ }\bibfield  {title} {\enquote {\bibinfo {title}
  {Approaching a {{Minimal Topological Electronic Structure}} in
  {{Antiferromagnetic Topological Insulator MnBi2Te4}} via {{Surface
  Modification}}}}, }\href {\doibase 10.1021/acs.nanolett.1c04930} {\bibfield
  {journal} {\bibinfo  {journal} {Nano Lett.}\ }\textbf {\bibinfo {volume}
  {22}},\ \bibinfo {pages} {4307} (\bibinfo {year} {2022})}\BibitemShut
  {NoStop}%
\bibitem [{\citenamefont {Lei}\ \emph {et~al.}(2022)\citenamefont {Lei},
  \citenamefont {Zhou}, \citenamefont {Hao}, \citenamefont {Liu}, \citenamefont
  {Yang}, \citenamefont {Sun}, \citenamefont {Ma}, \citenamefont {Ma},
  \citenamefont {Wang}, \citenamefont {Lu}, \citenamefont {Mei}, \citenamefont
  {Wang},\ and\ \citenamefont {He}}]{Lei22PRBMagnetically}%
  \BibitemOpen
  \bibfield  {author} {\bibinfo {author} {\bibfnamefont {X.}~\bibnamefont
  {Lei}},  \emph {et~al.},\ }\bibfield  {title} {\enquote {\bibinfo {title}
  {Magnetically tunable {Shubnikov--de Haas} oscillations in
  {MnBi$_2$Te$_4$}}}, }\href {\doibase 10.1103/PhysRevB.105.155402} {\bibfield
  {journal} {\bibinfo  {journal} {Phys. Rev. B}\ }\textbf {\bibinfo {volume}
  {105}},\ \bibinfo {pages} {155402} (\bibinfo {year} {2022})}\BibitemShut
  {NoStop}%
\bibitem [{\citenamefont {Zhang}\ \emph {et~al.}(2009)\citenamefont {Zhang},
  \citenamefont {Liu}, \citenamefont {Qi}, \citenamefont {Dai}, \citenamefont
  {Fang},\ and\ \citenamefont {Zhang}}]{Zhang09NPTopological}%
  \BibitemOpen
  \bibfield  {author} {\bibinfo {author} {\bibfnamefont {H.}~\bibnamefont
  {Zhang}}, \bibinfo {author} {\bibfnamefont {C.-X.}\ \bibnamefont {Liu}},
  \bibinfo {author} {\bibfnamefont {X.-L.}\ \bibnamefont {Qi}}, \bibinfo
  {author} {\bibfnamefont {X.}~\bibnamefont {Dai}}, \bibinfo {author}
  {\bibfnamefont {Z.}~\bibnamefont {Fang}}, \ and\ \bibinfo {author}
  {\bibfnamefont {S.-C.}\ \bibnamefont {Zhang}},\ }\bibfield  {title} {\enquote
  {\bibinfo {title} {Topological insulators in {{Bi$_2$Se$_3$}},
  {{Bi$_2$Te$_3$}} and {{Sb$_2$Te$_3$}} with a single {{Dirac}} cone on the
  surface}}, }\href {\doibase 10.1038/nphys1270} {\bibfield  {journal}
  {\bibinfo  {journal} {Nat. Phys.}\ }\textbf {\bibinfo {volume} {5}},\
  \bibinfo {pages} {438} (\bibinfo {year} {2009})}\BibitemShut {NoStop}%
\bibitem [{\citenamefont {Sun}\ \emph {et~al.}(2020)\citenamefont {Sun},
  \citenamefont {Wang}, \citenamefont {Zhang}, \citenamefont {Chen},
  \citenamefont {Zhao}, \citenamefont {Liu}, \citenamefont {Liu}, \citenamefont
  {Chen}, \citenamefont {Lu},\ and\ \citenamefont {Xie}}]{Sun20PRBAnalytical}%
  \BibitemOpen
  \bibfield  {author} {\bibinfo {author} {\bibfnamefont {H.-P.}\ \bibnamefont
  {Sun}}, \bibinfo {author} {\bibfnamefont {C.~M.}\ \bibnamefont {Wang}},
  \bibinfo {author} {\bibfnamefont {S.-B.}\ \bibnamefont {Zhang}}, \bibinfo
  {author} {\bibfnamefont {R.}~\bibnamefont {Chen}}, \bibinfo {author}
  {\bibfnamefont {Y.}~\bibnamefont {Zhao}}, \bibinfo {author} {\bibfnamefont
  {C.}~\bibnamefont {Liu}}, \bibinfo {author} {\bibfnamefont {Q.}~\bibnamefont
  {Liu}}, \bibinfo {author} {\bibfnamefont {C.}~\bibnamefont {Chen}}, \bibinfo
  {author} {\bibfnamefont {H.-Z.}\ \bibnamefont {Lu}}, \ and\ \bibinfo {author}
  {\bibfnamefont {X.~C.}\ \bibnamefont {Xie}},\ }\bibfield  {title} {\enquote
  {\bibinfo {title} {Analytical solution for the surface states of the
  antiferromagnetic topological insulator {{MnBi}}$_2${{Te}}$_4$}}, }\href
  {\doibase 10.1103/PhysRevB.102.241406} {\bibfield  {journal} {\bibinfo
  {journal} {Phys. Rev. B}\ }\textbf {\bibinfo {volume} {102}},\ \bibinfo
  {pages} {241406} (\bibinfo {year} {2020})}\BibitemShut {NoStop}%
\bibitem [{\citenamefont {Lu}\ \emph {et~al.}(2013)\citenamefont {Lu},
  \citenamefont {Zhao},\ and\ \citenamefont {Shen}}]{Lu13PRLQuantum}%
  \BibitemOpen
  \bibfield  {author} {\bibinfo {author} {\bibfnamefont {H.-Z.}\ \bibnamefont
  {Lu}}, \bibinfo {author} {\bibfnamefont {A.}~\bibnamefont {Zhao}}, \ and\
  \bibinfo {author} {\bibfnamefont {S.-Q.}\ \bibnamefont {Shen}},\ }\bibfield
  {title} {\enquote {\bibinfo {title} {Quantum {{Transport}} in {{Magnetic
  Topological Insulator Thin Films}}}}, }\href {\doibase
  10.1103/PhysRevLett.111.146802} {\bibfield  {journal} {\bibinfo  {journal}
  {Phys. Rev. Lett.}\ }\textbf {\bibinfo {volume} {111}},\ \bibinfo {pages}
  {146802} (\bibinfo {year} {2013})}\BibitemShut {NoStop}%
\bibitem [{\citenamefont {Shan}\ \emph {et~al.}(2010)\citenamefont {Shan},
  \citenamefont {Lu},\ and\ \citenamefont {Shen}}]{Shan10NJPEffective}%
  \BibitemOpen
  \bibfield  {author} {\bibinfo {author} {\bibfnamefont {W.-Y.}\ \bibnamefont
  {Shan}}, \bibinfo {author} {\bibfnamefont {H.-Z.}\ \bibnamefont {Lu}}, \ and\
  \bibinfo {author} {\bibfnamefont {S.-Q.}\ \bibnamefont {Shen}},\ }\bibfield
  {title} {\enquote {\bibinfo {title} {Effective continuous model for surface
  states and thin films of three-dimensional topological insulators}}, }\href
  {\doibase 10.1088/1367-2630/12/4/043048} {\bibfield  {journal} {\bibinfo
  {journal} {New J. Phys.}\ }\textbf {\bibinfo {volume} {12}},\ \bibinfo
  {pages} {043048} (\bibinfo {year} {2010})}\BibitemShut {NoStop}%
\bibitem [{\citenamefont {Lu}\ \emph {et~al.}(2010)\citenamefont {Lu},
  \citenamefont {Shan}, \citenamefont {Yao}, \citenamefont {Niu},\ and\
  \citenamefont {Shen}}]{Lu10PRBMassive}%
  \BibitemOpen
  \bibfield  {author} {\bibinfo {author} {\bibfnamefont {H.-Z.}\ \bibnamefont
  {Lu}}, \bibinfo {author} {\bibfnamefont {W.-Y.}\ \bibnamefont {Shan}},
  \bibinfo {author} {\bibfnamefont {W.}~\bibnamefont {Yao}}, \bibinfo {author}
  {\bibfnamefont {Q.}~\bibnamefont {Niu}}, \ and\ \bibinfo {author}
  {\bibfnamefont {S.-Q.}\ \bibnamefont {Shen}},\ }\bibfield  {title} {\enquote
  {\bibinfo {title} {Massive {{Dirac}} fermions and spin physics in an
  ultrathin film of topological insulator}}, }\href
  {https://link.aps.org/doi/10.1103/PhysRevB.81.115407} {\bibfield  {journal}
  {\bibinfo  {journal} {Phys. Rev. B}\ }\textbf {\bibinfo {volume} {81}},\
  \bibinfo {pages} {115407} (\bibinfo {year} {2010})}\BibitemShut {NoStop}%
\bibitem [{Par()}]{Parameters_transistors}%
  \BibitemOpen
  \href@noop {} {\bibinfo  {journal} {Numerical transport calculations are
  based on the discretized 3D bulk Hamiltonian on a cubic lattice. The bulk
  parameters are adopted to the material MnBi$_2$Te$_4$ as $A_1$=270.23
  meV${\cdot}$nm, $A_2$=319.64 meV${\cdot}$nm, $B_1$=-119.05
  meV${\cdot}$nm$^2$, $B_2$=-94.05 meV${\cdot}$nm$^2$, $M_0$=-116.50 meV,
  $m$=50 meV. Here, we choose the parameter $V_z=5$ meV in
  Figs.~\ref{Fig:AFM_Normal_conductance}(a),~\ref{Fig:AFM_Normal_conductance}(b)
  and~\ref{Fig:AFM_Normal_conductance}(d) and
  Figs.~\ref{Fig:Conductance_Nx_var}(a) and~\ref{Fig:Conductance_Nx_var}(b),
  and $V_z=3$ meV in Fig.~\ref{Fig:AFM_Normal_conductance}(c). The system size
  is $N_x=20$ and $N_y=20$ unless speciﬁed}\ }\BibitemShut {NoStop}%
\bibitem [{\citenamefont {Landauer}(1970)}]{landauer1970electrical}%
  \BibitemOpen
\bibfield  {journal} {  }\bibfield  {author} {\bibinfo {author} {\bibfnamefont
  {R.}~\bibnamefont {Landauer}},\ }\bibfield  {title} {\enquote {\bibinfo
  {title} {Electrical resistance of disordered one-dimensional lattices}},
  }\href {\doibase 10.1080/14786437008238472} {\bibfield  {journal} {\bibinfo
  {journal} {The Philosophical Magazine}\ }\textbf {\bibinfo {volume} {21}},\
  \bibinfo {pages} {863} (\bibinfo {year} {1970})}\BibitemShut {NoStop}%
\bibitem [{\citenamefont {B\"uttiker}(1986)}]{buttiker1986four}%
  \BibitemOpen
  \bibfield  {author} {\bibinfo {author} {\bibfnamefont {M.}~\bibnamefont
  {B\"uttiker}},\ }\bibfield  {title} {\enquote {\bibinfo {title}
  {{Four-Terminal Phase-Coherent Conductance}}}, }\href
  {https://link.aps.org/doi/10.1103/PhysRevLett.57.1761} {\bibfield  {journal}
  {\bibinfo  {journal} {Phys. Rev. Lett.}\ }\textbf {\bibinfo {volume} {57}},\
  \bibinfo {pages} {1761} (\bibinfo {year} {1986})}\BibitemShut {NoStop}%
\bibitem [{\citenamefont {Datta}(1997)}]{Datta97Electronic}%
  \BibitemOpen
  \bibfield  {author} {\bibinfo {author} {\bibfnamefont {S.}~\bibnamefont
  {Datta}},\ }\href@noop {} {\emph {\bibinfo {title} {Electronic {{Transport}}
  in {{Mesoscopic Systems}}}}}\ (\bibinfo  {publisher} {{Cambridge University
  Press}},\ \bibinfo {year} {1997})\BibitemShut {NoStop}%
\bibitem [{\citenamefont {MacKinnon}(1985)}]{MacKinnon85ZPB-CMCalculation}%
  \BibitemOpen
  \bibfield  {author} {\bibinfo {author} {\bibfnamefont {A.}~\bibnamefont
  {MacKinnon}},\ }\bibfield  {title} {\enquote {\bibinfo {title} {The
  calculation of transport properties and density of states of disordered
  solids}}, }\href {\doibase 10.1007/BF01328846} {\bibfield  {journal}
  {\bibinfo  {journal} {Z. Physik B - Condensed Matter}\ }\textbf {\bibinfo
  {volume} {59}},\ \bibinfo {pages} {385} (\bibinfo {year} {1985})}\BibitemShut
  {NoStop}%
\bibitem [{\citenamefont {Zeugner}\ \emph {et~al.}(2019)\citenamefont
  {Zeugner}, \citenamefont {Nietschke}, \citenamefont {Wolter}, \citenamefont
  {Ga{\ss}}, \citenamefont {Vidal}, \citenamefont {Peixoto}, \citenamefont
  {Pohl}, \citenamefont {Damm}, \citenamefont {Lubk}, \citenamefont {Hentrich},
  \citenamefont {Moser}, \citenamefont {Fornari}, \citenamefont {Min},
  \citenamefont {Schatz}, \citenamefont {Ki{\ss}ner}, \citenamefont
  {{\"U}nzelmann}, \citenamefont {Kaiser}, \citenamefont {Scaravaggi},
  \citenamefont {Rellinghaus}, \citenamefont {Nielsch}, \citenamefont {Hess},
  \citenamefont {B{\"u}chner}, \citenamefont {Reinert}, \citenamefont
  {Bentmann}, \citenamefont {Oeckler}, \citenamefont {Doert}, \citenamefont
  {Ruck},\ and\ \citenamefont {Isaeva}}]{Zeugner19CMChemical}%
  \BibitemOpen
  \bibfield  {author} {\bibinfo {author} {\bibfnamefont {A.}~\bibnamefont
  {Zeugner}},  \emph {et~al.},\ }\bibfield  {title} {\enquote {\bibinfo {title}
  {Chemical {{Aspects}} of the {{Candidate Antiferromagnetic Topological
  Insulator} {MnBi}$_{2}${Te}$_{4}$}}}, }\href
  {https://doi.org/10.1021/acs.chemmater.8b05017} {\bibfield  {journal}
  {\bibinfo  {journal} {Chem. Mater.}\ }\textbf {\bibinfo {volume} {31}},\
  \bibinfo {pages} {2795} (\bibinfo {year} {2019})}\BibitemShut {NoStop}%
\bibitem [{\citenamefont {Shikin}\ \emph {et~al.}(2021)\citenamefont {Shikin},
  \citenamefont {Estyunin}, \citenamefont {Zaitsev}, \citenamefont {Glazkova},
  \citenamefont {Klimovskikh}, \citenamefont {Filnov}, \citenamefont {Rybkin},
  \citenamefont {Schwier}, \citenamefont {Kumar}, \citenamefont {Kimura},
  \citenamefont {Mamedov}, \citenamefont {Aliev}, \citenamefont {Babanly},
  \citenamefont {Kokh}, \citenamefont {Tereshchenko}, \citenamefont {Otrokov},
  \citenamefont {Chulkov}, \citenamefont {Zvezdin},\ and\ \citenamefont
  {Zvezdin}}]{Shikin21PRBSampledependent}%
  \BibitemOpen
  \bibfield  {author} {\bibinfo {author} {\bibfnamefont {A.~M.}\ \bibnamefont
  {Shikin}},  \emph {et~al.},\ }\bibfield  {title} {\enquote {\bibinfo {title}
  {Sample-dependent {{Dirac-point}} gap in {MnBi}$_{2}${Te}$_{4}$ and its
  response to applied surface charge: {{A}} combined photoemission and
  \textit{ab initio} study}}, }\href {\doibase 10.1103/PhysRevB.104.115168}
  {\bibfield  {journal} {\bibinfo  {journal} {Phys. Rev. B}\ }\textbf {\bibinfo
  {volume} {104}},\ \bibinfo {pages} {115168} (\bibinfo {year}
  {2021})}\BibitemShut {NoStop}%
\bibitem [{\citenamefont {Garnica}\ \emph {et~al.}(2022)\citenamefont
  {Garnica}, \citenamefont {Otrokov}, \citenamefont {Aguilar}, \citenamefont
  {Klimovskikh}, \citenamefont {Estyunin}, \citenamefont {Aliev}, \citenamefont
  {Amiraslanov}, \citenamefont {Abdullayev}, \citenamefont {Zverev},
  \citenamefont {Babanly}, \citenamefont {Mamedov}, \citenamefont {Shikin},
  \citenamefont {Arnau}, \citenamefont {{de Parga}}, \citenamefont {Chulkov},\
  and\ \citenamefont {Miranda}}]{Garnica22nQMNative}%
  \BibitemOpen
  \bibfield  {author} {\bibinfo {author} {\bibfnamefont {M.}~\bibnamefont
  {Garnica}},  \emph {et~al.},\ }\bibfield  {title} {\enquote {\bibinfo {title}
  {Native point defects and their implications for the {{Dirac}} point gap at
  {MnBi}$_{2}${Te}$_{4}$ (0001)}}, }\href {\doibase 10.1038/s41535-021-00414-6}
  {\bibfield  {journal} {\bibinfo  {journal} {npj Quantum Mater.}\ }\textbf
  {\bibinfo {volume} {7}},\ \bibinfo {pages} {1} (\bibinfo {year}
  {2022})}\BibitemShut {NoStop}%
\end{thebibliography}
%

\end{document}